\documentclass[journal]{IEEEtran}
\usepackage{textcomp}
\usepackage{xcolor}
\usepackage{comment}
\usepackage{capt-of}
\usepackage{listings}
\usepackage{balance}
\usepackage{algorithm,algpseudocode}
\usepackage{enumitem}
\usepackage{threeparttable}
\usepackage{multirow}
\usepackage{multicol}
\usepackage{pifont}
\usepackage{caption}
\usepackage{subcaption}
\usepackage{color}
\usepackage{framed}
\usepackage{booktabs}
\usepackage{graphicx}
\usepackage{amsmath,amssymb,amsfonts}
\usepackage{dirtytalk}
\usepackage{hyperref}

\newcounter{t0d0_counter}
\newcommand{\nofixme}[1]{
}

\ifCLASSINFOpdf
\else
\fi
\begin{document}

%
\title{OSRM-CCTV: Open-source CCTV-aware \\ routing and navigation system \\ for privacy, anonymity and safety (Preprint)}

%
%
%

\author{Lauri~Sintonen, 
        Hannu~Turtiainen, 
        Andrei~Costin, 
        Timo~H\"{a}m\"{a}l\"{a}inen
        and~Tuomo~Lahtinen
\thanks{
L. Sintonen, H. Turtainen, \textbf{A. Costin \emph{(Corresponding and original idea's author)}}, T. H\"{a}m\"{a}l\"{a}inen T. Lahtinen are with the 
Department of Information Technology, University of Jyv\"{a}skyl\"{a}, P.O. Box 35, Jyv\"{a}skyl\"{a}, 40014 Finland 
e-mail: 
\href{mailto:lauri.m.j.sintonen@student.jyu.fi}{lauri.m.j.sintonen@student.jyu.fi}, 
\href{mailto:turthzu@jyu.fi}{turthzu@jyu.fi}, 
\href{mailto:ancostin@jyu.fi}{ancostin@jyu.fi},
\href{mailto:timoh@jyu.fi}{timoh@jyu.fi},
\href{mailto:tuomo.t.lahtinen@student.jyu.fi}{tuomo.t.lahtinen@student.jyu.fi} \\ 
}
}

%
%

\markboth{OSRM-CCTV: Open-source CCTV-aware routing and navigation system for privacy, anonymity and safety (Preprint)}%
{Sintonen \MakeLowercase{\textit{et al.}}}

%



\maketitle

\begin{abstract}

For the last several decades, the increased, widespread, unwarranted, and unaccountable 
use of Closed-Circuit TeleVision (CCTV) cameras globally has raised concerns about privacy risks. 
Additional recent features of many CCTV cameras, such as Internet of Things (IoT) 
connectivity and Artificial Intelligence (AI)-based facial recognition, only increase concerns among privacy advocates. 
Therefore, on par \emph{CCTV-aware solutions} must exist that provide privacy, safety, and cybersecurity features.
We argue that an important step forward is to develop solutions addressing privacy concerns via routing and navigation systems (e.g., OpenStreetMap, Google Maps) that provide 
both privacy and safety options for areas where cameras are known to be present. 
However, at present no routing and navigation system, whether online or offline, provide corresponding CCTV-aware functionality. 

In this paper we introduce \emph{OSRM-CCTV} -- the first and only CCTV-aware routing and navigation system 
designed and built for privacy, anonymity and safety applications.
We validate and demonstrate the effectiveness and usability of the system on a handful of 
synthetic and real-world examples.
To help validate our work as well as to further encourage the development and wide adoption of the system, 
we release \emph{OSRM-CCTV} as open-source. 

\end{abstract}


\IEEEpeerreviewmaketitle


\section{Introduction}
\label{sec:intro}

\IEEEPARstart{I}{n} the modern world, public spaces of many cities are being surveilled by close-circuit television (CCTV) cameras to a considerable extent. It is estimated that globally there are around 770 million CCTV cameras in use and that their amount could rise up to one billion during the year 2021~\cite{bischoff-2021,cnbc2019billion}. As an example, there are approximately half a million CCTV cameras in London and an average person living there is recorded on camera 300 times everyday~\cite{caughtoncamera2019cctvlond}. In the United States, it is likely that people get recorded by a CCTV camera over fifty times per day~\cite{ipvm2016us}. In 2019, a person documented 49 CCTV cameras on the way to work in New York City~\cite{bi2019cctv} and described it dystopian. 

The discourse on CCTV surveillance has ethical dimensions. Von Hirsch argues that CCTV surveillance is sometimes covert and that often people believe that they are not under CCTV surveillance when they actually are~\cite{vonhirsch2000ethics}. Furthermore, according to a 2016 survey, an average citizen of the United States assumed being recorded by four or less CCTV cameras per day while the real figure is likely over 10 times larger~\cite{ipvm2016us}. Considering the amount of CCTV cameras having been installed globally and the fact that people can be detected and recorded by them, adding face recognition to the pattern opens up an unsettling possibility to also automatically identify people by CCTV cameras~\cite{hu2004survey,wheeler2010face,axis-id-rec}. Overall, there is controversy about the ethics of CCTV surveillance and whole books have been written on the subject~\cite{vonhirsch2000ethics,larsen2011setting}. Moreover, the CCTV cameras, Digital Video Recorders (DVR) and Video Surveillance Systems (VSS) are notoriously known to be vulnerable to cybersecurity attacks and hacks~\cite{costin2016security}, therefore it is a reasonable to assume that the CCTV cameras overlooking public places and routes may be at any time under the control of unauthorized persons hence posing a direct threat to privacy. 

In this context, we argue that it is important to create \emph{CCTV-aware} solutions and technologies that allow people at least a slight option to choose between being under CCTV surveillance (or not!) whenever they walk, cycle or drive in public spaces. We approach the question from the perspective of routing and navigation. While there is a substantial amount of existing routing solutions and technologies~\cite{osm-route-online,osm-route-offline,osm-based,luxen2011real,bast2016route,szczerba2000robust}, to the best of our knowledge none of them focuses on the issue at hand, to provide a \emph{CCTV-aware} routing solution. 

In this paper, we propose one such \emph{CCTV-aware} solution. Our \emph{CCTV-aware} solution enables two modes of routing. Firstly, \emph{privacy-mode} which aids maintaining privacy by choosing a route where CCTV cameras are avoided. \emph{Privacy-mode} is strict in the sense that it will never choose a route that leads through the field of vision of a CCTV camera.  Avoiding CCTV cameras would be desirable anytime when privacy or anonymity is important. Secondly, \emph{safety-mode} which aids maintaining safety by choosing a weighed route that leads through the fields of vision of CCTV cameras. \emph{Safety-mode} is called safe because sometimes (e.g., at night and for the preference of staying physically safe) one would prefer to be actually detected and recorded by CCTV cameras. 

It is worth to mention that neither of these routing modes are guaranteed to completely maintain either privacy or safety as described in this paper. This is mainly because of the lack of up-to-date and accurate data about the CCTV cameras. The relevant data in question would contain information about for example CCTV camera model and lense, the radius, the angle of view and direction of the CCTV cameras, and data like this is more often than not is unavailable. Currently, to our knowledge there is no automated way to gather data about CCTV cameras to be used in our \emph{CCTV-aware} approach. However, regarding the collection of relevant data we have also developed and demonstrated the first Computer Vision (CV) models that can be used to detect CCTV cameras in images (e.g., street-level/street view, indoors). For example, our CV models achieve an accuracy of up to 98.7\%~\cite{turtiainen2020towards} and we are currently in the process of mapping large areas around the globe. 

The rest of the paper is organized as follows. 
In Section~\ref{sec:related} we present earlier work that has been done to allow different types of routing. 
Then, in Section~\ref{sec:impl} we provide an outline of how we have implemented our \emph{CCTV-aware} solution. 
Next, in Section~\ref{sec:eval} we describe how our \emph{CCTV-aware} solution functions where we present different examples of both \emph{privacy-mode} and \emph{security-mode} on a synthetic evaluation map where different routing scenarios can be demonstrated. 
We present examples based on real data in Section~\ref{sec:realworld}. 
We conclude this paper with Section~\ref{sec:concl}.

Our main contributions with this work are:
\begin{enumerate}
\item We are the first (to the best of our knowledge) to propose, implement, and evaluate CCTV-aware routing for privacy, anonymity and safety applications.
\item We demonstrate the usefulness and the feasibility of CCTV-aware routing applied to both synthetic and real-world examples.
\item We release (upon peer-review acceptance) the relevant artifacts (e.g., code, data, documentation) as open-source: \url{https://github.com/Fuziih}
\end{enumerate}


\section{Related work}
\label{sec:related}

\emph{OpenStreetMap (OSM)} is a collaborative open-source project, which aims to supply free and editable maps of the world. Many reputable projects~\cite{osm-route-online,osm-route-offline,osm-based,luxen2011real,bast2016route,szczerba2000robust} (e.g., routing solutions and technologies for different use-cases) base their work on the OSM. Most routing solutions focus on standard driving, cycling, and walking schemes. In addition to this, there are some less common/popular yet useful solutions, such as finding wheel-chair accessible routes~\cite{openrouteservice,wheelchair,crowdsourcing-mobility-impaired}, finding curvy roads~\cite{curvature}, and solutions that calculate routes based on the day arc~\cite{shadow-as-route-2016,keithma2018parasol,deilami2020allowing}.

Recently, Siriaraya et al.~\cite{quality-aware} argued that in the modern contemporary society, pedestrians prefer to set ``alternative routing criteria'' rather than the more common routing criteria such as duration and length. The authors refer to this as ``routing based on different qualities``, i.e., \emph{qualitative-aware routing}. The authors identified five routing quality categories: safety, well-being, effort, exploration, and pleasure. Correspondingly to our work, safety category is acknowledged by the authors as the need for avoiding areas with a high occurrence of past criminal activities. Although privacy is not mentioned nor categorized, it could be accommodated into the safety category of~\cite{quality-aware}.

When pursuing to achieve a \emph{CCTV-aware} routing solution, inspiration could be drawn from the day arc routing solutions. For example, Olaverri Monreal et al.~\cite{shadow-as-route-2016} created sun-avoiding routing solution and Ma~\cite{keithma2018parasol, parasol2018github} developed a solution to either avoid or face the sun during travel. Recently, Deilami
et al.~\cite{deilami2020allowing} provided a routing solution, which takes tree shading into consideration.
The routing solutions in these works could be used to establish our \emph{CCTV-aware} routing, however, there are a few pitfalls with that approach. Firstly, the data requirements for these solutions are quite specific and reaching them could prove to be problematic. Secondly, as an \emph{OSM} ``Way`` object is defined as a collection of nodes, a line, and the width is saved only as a key–value tag~\cite{osm-way,osm-creaking}, this fundamental issue would need to be addressed.

Privacy and safety routing has been previously researched to a certain, yet limited, degree. Bao et al.~\cite{safe-lighting} proposed a routing solution with multiple safe pedestrian routing conditions. Hirozaku et al.~\cite{street-illumination} took into account street lights for safer walking routes. Keler and Mazimpaka~\cite{safety-aware-routing} employed crowd-sourcing in their efforts to create safe routing. Lastly, Tessio et al.~\cite{customized-pedestrian-routes} were set on finding ways for users to find routing through green areas, social places and quieter streets. Lahtinen et al.~\cite{lahtinen2020feasibility} investigated just the initial feasibility of CCTV-aware routing solutions and the quantitative impact on the routing options. 

However, none of the previous works address the privacy, anonymity and safety routing approaches given the exponential growth of CCTV camera and video surveillance presence in public spaces and routes~\cite{bi2019cctv}.
With this work we try to close this gap as it is also part of our bigger vision revolving around CCTV cameras, video surveillance, digital privacy and anonymity. Therefore, we present that related work as well. 
First, we designed a Computer Vision (CV) model for detecting CCTV and video surveillance cameras in images and video frames~\cite{turtiainen2020towards}. 
Second, in order to improve and accelerate the training of our CV model~\cite{turtiainen2020towards}, we created from scratch a novel image annotation tool \emph{BRIMA}~\cite{lahtinen2021brima} which is a browser extension that provides an effortless way to map CCTV cameras using services such as Google Maps Street View~\cite{google-maps}, Yandex~\cite{yandex-maps} or Baidu~\cite{baidu-maps}. 
Third, we published an initial feasibility-study work where we mapped 450 CCTV cameras in the city of Jyv\"askyl\"a, Finland, and conducted routing experiments concluding that our system is feasible for further exploration and development, as it supports both the \emph{privacy-first} and \emph{safety-first} approaches~\cite{lahtinen2021towards}.

\section{Implementation details}
\label{sec:impl}

This section contains a description of the most relevant tools and libraries as well as the data manipulation process that are used to enable \emph{CCTV-aware} routing with two alternative modes called \emph{privacy-mode} and \emph{safety-mode}. The routing process is two-fold. First, \emph{OSM} data is processed so that the resulting \emph{OSM} file contains 1) the CCTV cameras as nodes, 2) entrance and exit nodes of the areas that intersect with the fields of vision of CCTV cameras, and 3) ways that are not merely lines without a width but ones that have an actual width on the map.  \emph{OSM} Way objects do not have an actual width as they are defined as a collection of nodes, and the width is saved only as a key–value tag~\cite{osm-way,osm-creaking}. Having an actual width is essential in enabling \emph{CCTV-aware} routing, because if a CCTV camera surveys only part of one side of a way, one should still be able to traverse the other side of that way, and width is needed to allow this possibility. Second, our \emph{OSRM-CCTV} routing implementation itself is based on \emph{Open Source Routing Machine (OSRM)}~\cite{luxen-vetter-2011, osrm-first} using a custom \emph{OSM} file created in the processing phase. \emph{Privacy-mode} and \emph{safety-mode} are described in \emph{Lua}~\cite{lua-about} programming language as customized profiles that the standard \emph{OSRM} supports by default~\cite{osrm-profiles}.

\subsection{Tools and Libraries}

In this section, the most relevant tools and libraries used for data processing are described. The effort enabling \emph{CCTV-aware} routing is heavily weighed on processing \emph{OSM} data and the programming language used for that is \emph{Python}~\cite{python}. The relevant tools and libraries are \emph{Osmium}~\cite{osmium-osmcode} and its \emph{Python} bindings, \emph{PyOsmium}~\cite{pyosmium-osmcode}, \emph{Shapely}~\cite{shapely} and \emph{PyProj}~\cite{pyproj}. We run the backend using \emph{Docker} and \emph{Docker-Compose}~\cite{docker} on an \emph{Ubuntu 18.04}~\cite{ubuntu} platform. Internal routing between \emph{Docker} containers is performed with \emph{Nginx}~\cite{nginx}.

\emph{Osmium} is a \textit{``multipurpose command line tool for working with OpenStreetMap data based on the Osmium library''}~\cite{osmium-osmcode}. In our work, \emph{PyOsmium}~\cite{pyosmium-osmcode}, the \emph{Python} bindings of the \emph{Osmium} library, is the main library used in adding cameras and splitting ways in \emph{OSM} data, more specifically, \emph{OSM XML} data~\cite{osm-xml}. \emph{PyOsmium} is also used to write the resulting \emph{OSM} file that is used in routing with \emph{Open Source Routing Machine}. \emph{Shapely} is a \textit{``Python package for manipulation and analysis of planar geometric objects''}~\cite{shapely}. It is used to transform \emph{OSM} data to a more modifiable form to perform most of the necessary geometric transformations and calculations that are necessary in manipulating \emph{OSM} data. Since Shapely assumes data on a Cartesian plane, we use \emph{Pyproj}~\cite{pyproj} for spatial projections.

\subsection{Data Processing}

In this section, the data processing phase of our \emph{CCTV-aware} routing solution is described. This involves the necessary steps in order to enable \emph{CCTV-aware} routing with \emph{OSRM-CCTV}, and can be divided largely into two parts:
\begin{enumerate}
    \item manipulating OSM file to contain CCTV data
    \item preparing final OSM data for routing with \emph{OSRM-CCTV}
\end{enumerate}

\subsubsection{Manipulating OSM data}

\begin{algorithm}
\begin{algorithmic}[1]
\caption{OSM data manipulation.}
\label{alg:1}
\Procedure{main}{$data$}:
    \ForAll {line in data}
        \State $coords$ = line["location of camera"]
        \State $cam\_type$ = line["directed"] OR line["round"] 
        \State $radius$ = line["how far the camera sees"]
        \State $angle$ = line["how wide the camera sees"]
        \State $direction$ = line["camera pointing direction"]
        \State \textbf {call}
        $handle\_surveillance\_area(coords,cam\_type,\linebreak \indent \indent radius,angle,direction,writer,osm)$
    \EndFor
\State compile\_data\_and\_add\_new\_and\_old\_ways\_to\_new\_osm
\EndProcedure
\newline

\Procedure{handle\_surveillance\_area}{$coords$, $cam\_type$, $radius$, $angle$, $direction$, $writer$, $osm$}:
    \State writer.add\_camera(coords)
    \State determine\_surveillance\_area(cam\_type)
    \State extract\_area = determine\_extract\_area(coords)
    \State \textbf {call} $create\_tmp\_extract(extract\_area,osm\_file)$
    \ForAll {travellable\_way in tmp\_extract.ways}
        \State add\_split\_ways
        \State gather\_data\_for\_compilation
        \State writer.add\_way\_nodes
        \State \textbf {call} handle\_intersections\_aov\_way
        \State gather\_data\_for\_compilation
    \EndFor 
\EndProcedure
\newline

\Procedure{create\_tmp\_extract}{$extract\_area$, $osm\_file$}:
    \State save\_tmp\_extract\_on\_disk(from osm\_file)
    \State split\_ways\_in\_tmp\_extract \Comment use split\_ways.py
    \ForAll {way in travellable\_ways}
        \State create\_parallel\_ways
    \EndFor
\EndProcedure
\newline

    \Procedure{handle\_intersections\_aov\_way}{}:
   \If{intersects}:
       \State add\_entrance\_and\_exit\_nodes\_to\_cam\_aov\_boundary
       \State create\_weights\_for\_safety\_mode
       \State gather\_data\_for\_compilation
   \EndIf
   \State gather\_data\_of\_split\_ways\_for\_compilation
   \State gather\_data\_of\_regular\_ways\_for\_compilation
\EndProcedure
\end{algorithmic}
\end{algorithm}

Algorithm~\ref{alg:1} describes the most important aspects of the \emph{OSM)} data manipulation phase in pseudocode. It is simplified and not a completely accurate representation of everything that happens in the code. In essence, the algorithm uses an existing \emph{OSM XML}\cite{osm-xml} file and writes a new one based on a Comma-Separated Value (CSV) file which includes the properties of the cameras that are to be added to the resulting \emph{OSM XML} file. These properties are GPS coordinates (e.g., latitude, longitude), type, radius, angle and direction of vision of each new camera. Two types for the cameras are defined: \emph{directed} and \emph{round}. A \emph{round} camera has a 360 degree angle of view, while an angle narrower than that would represent a \emph{directed} camera type. \emph{Radius} means how far the camera can ``see'' for its particular function, for example, ``detect faces'', ``recognize faces``, or ``recognize license-plate numbers'' (i.e., depending on privacy/safety-impact scenario being modeled and evaluated). The radius can be chosen according to case-specific needs (e.g., depending on characteristics of CCTV cameras and lenses), and its exact value is not relevant for performance evaluations in this paper. \emph{Angle}, or horizontal Field of View (FoV) in our case, describes how wide the angle of the camera vision is, hence, it also affects the determination of the type of the camera between \emph{round} and \emph{directed}. \emph{Direction} means the direction to which the center of the field of vision of the camera is pointing.

The resulting \emph{OSM XML} file will include the new CCTV cameras as nodes.  Also, the intersection points of routable ways and the fields of vision of the CCTV cameras are represented as nodes on the ways. These nodes are marked with an access tag that points to surveillance. From the perspective of \emph{CCTV-aware} routing, the nodes with an access tag are enough in determining when one would enter the field of vision of a CCTV camera.  Depending on the selected routing mode, the router would either avoid the access tags that point towards surveillance or pursue routing through these nodes. Finally, the ways in the resulting \emph{OSM XML} are split in three whenever there is a CCTV camera nearby. The three split ways represent the left, the middle and the right side of the way.

Next, some clarification is provided for some parts in the pseudocode Algorithm~\ref{alg:1}. In \emph{PyOsmium}~\cite{pyosmium-osmcode}, the Python bindings of the \emph{Osmium} library~\cite{osmium-osmcode}, \emph{writer} is used for writing \emph{OSM} objects on the resulting \emph{OSM} file. Hence, when \emph{writer} is mentioned, something is written on that file. The \emph{osm} parameter seen for instance on line 12 references to the original \emph{OSM} file that has no added CCTV cameras. 

Based on the type, the angle and the radius of the field of vision of the CCTV camera, the subroutine \emph{determine\_surveillance\_area} is used for determining the area that the camera surveys. When the camera type is "round" and the angle of the camera vision is 360 degrees, the surveilled area is circular. In the case of a directed CCTV camera, the surveilled area is a sector of a circle. In this subroutine, \emph{determine\_extract\_area} determines the area which is extracted from the original \emph{OSM} file. This area is a circle with a diameter of 50. In \emph{create\_tmp\_extract}, the determined extract area is used to execute the extraction and writing a temporary OSM file on the disk. Also, all travellable ways that are found inside the circle are split in two using another script, \emph{split\_ways.py}.  Splitting ways means creating two additional ways parallel to the original way.  The distance of each of the parallel ways will be the value of the width tag of the original way divided by 2 – creating a left and a right border of the way. A final temporary file is created. Data is manipulated in intermediary extracts to mitigate the memory-extensiveness of manipulating large \emph{OSM XML}~\cite{osm-xml} files.

Now, reaching the line 17 in Algorithm~\ref{alg:1}, all ways of possible travel in the newly-created temporary \emph{OSM} file are iterated. The split ways are added to the the final resulting \emph{OSM} file and the intersections with the ways and the borders of the CCTV surveillance areas that were determined earlier are handled in \emph{handle\_intersections\_aov\_way}. When a way penetrates a camera-surveilled area, one or two intersections occur. Entrance and exit nodes are added at the intersection points. Also, weights for \emph{OSRM-CCTV} traffic updates~\cite{osrm-traffic} used with \emph{safety-mode} are added and written on disk to a \emph{CSV} file.

Finally, in Algorithm~\ref{alg:1} there are mentions about gathering data. These mentions refer to every time that some data is saved into variables for the final compilation executed in the end of Algorithm~\ref{alg:1} on line 10, at \emph{compile\_data\_and\_add\_new\_and\_old\_ways\_to\_new\_osm}.

\subsubsection{Preparing data for routing with OSRM-CCTV}

The second part of the data processing is executed by \emph{OSRM-CCTV}, which involves parts for extracting, partitioning and customizing both the \emph{OSM} file (resulting from the first part of the data processing phase) and the profile file which determines which mode is used -- \emph{privacy} or \emph{safety}. Extracting, partitioning and customizing are performed to create all the necessary files for running \emph{OSRM-CCTV} router. 

\subsection{Privacy-mode and safety-mode}

Our \emph{CCTV-aware} routing solution has two alternative modes called \emph{privacy-mode} and \emph{safety-mode}. Their functioning utilizes the entrance and exit nodes on the ways that intersect with the fields of vision of CCTV cameras. Not allowing to traverse through these nodes and through the field of vision of a CCTV camera would maintain privacy, hence, \emph{privacy-mode} is being used. \emph{Privacy-mode} is implemented as a \emph{Lua} profile~\cite{osrm-profiles} by blacklisting access tags that indicate surveillance. \emph{Safety-mode} works vice versa. \emph{Safety-mode} does not blacklist the access tags in the \emph{Lua} profile, but it is instead enabled by using traffic updates~\cite{osrm-traffic} of the standard \emph{OSRM} functionality. These traffic updates are used to weigh the calculations so that the route will run through areas with more CCTV cameras. These calculations use weights that are added to the \emph{OSM} file in the data manipulation phase before running the router. 

\subsection{Running Open Street Routing Machine}

In this section, running \emph{OSRM-CCTV} is described. \emph{OSRM-CCTV} (as inherited from baseline \emph{OSRM} project) has a backend~\cite{osrm-backend} and a frontend~\cite{osrm-frontend}. We have chosen to use the default \emph{OSRM} backend (i.e., controlling the routing via blacklist tags, weights and traffic updates), but decided fork from the standard \emph{OSRM} frontend~\cite{osrm-frontend-fossgis} as it is more suitable for the needs for toggling between the \emph{privacy-mode} and \emph{safety-mode}. While not necessary functionality for the actual routing purposes, displaying the CCTV cameras on the map that is rendered by the \emph{OSRM-CCTV} frontend requires hosting its own tile server. For the tile server, we decided to use \emph{openstreetmap-tile-server} by Overvoorde~\cite{tile-server} with style modifications made by Townsend~\cite{someoneelse}. We run the services in virtual containers using \emph{Docker} and \emph{Docker Compose}~\cite{docker}. Internal routing between Docker containers is performed with \emph{Nginx}~\cite{nginx}.

Running \emph{OSRM-CCTV} and handling requests works as follows. After the data processing phase, two instances of the \emph{OSRM-CCTV} backend and the \emph{OSRM-CCTV} frontend are run within \emph{Docker} containers. One of the two backend containers is dedicated to \emph{privacy-mode} and the other one to \emph{safety-mode}. Two distinct containers are necessary because the standard \emph{OSRM} requires a static, pre-processed set of files for each different mode. A user can access the frontend with a web browser and input a typical query for requesting a route from point A to point B, also choosing either \emph{privacy-mode} or \emph{safety-mode} from a drop-down menu. The request is sent to a reverse proxy, \emph{Nginx}~\cite{nginx}, which uses the \emph{Uniform Resource Locator (URL)} to forward the request to the correct backend container, depending on the condition of either \emph{privacy-mode} or \emph{safety-mode} being used. The backend computes the route and sends it back to be drawn on a map by the frontend that is visible on the web browser.


\section{Evaluation}
\label{sec:eval}

In this section, we present the function of our \emph{CCTV-aware} routing solution on a synthetic map that was created for the evaluation. First, we provide screenshots from \emph{Java OpenStreetMap Editor (JOSM)}~\cite{josm} to illustrate what \emph{OSM} files without and with cameras added by our algorithm look like.
On the synthetic evaluation map, we first describe how the standard \emph{OSRM} works by default and without either \emph{privacy-mode} or \emph{safety-mode}. We then proceed into showing examples of \emph{privacy-mode} where the calculated route avoids CCTV cameras completely. Following \emph{privacy-mode}, we demonstrate \emph{safety-mode} in which the route is weighed to prefer surveilled routes. \emph{Safety-mode} is not as absolute as \emph{privacy-mode} as it does not choose too long routes even if they are surveilled by CCTV cameras. After demonstrating \emph{privacy-mode} and \emph{safety-mode} with a comprehensive synthetic setup (Section~\ref{sec:eval}), we finally show routing examples with real-world data (Section~\ref{sec:realworld}). 

\subsection{Java OpenStreetMap Editor}

In this section, we provide screenshots of \emph{OSM} data viewed in \emph{JOSM}. The figures illustrate what \emph{JOSM} in general and with CCTV cameras added with our script look like.  The directions and the radii of the cameras are random as our data does not yet contain that data. For clarity, the borders of the areas surveilled by cameras are marked by \emph{OSM} node objects. When actually using the router, these nodes should be omitted from the data, since they do not affect the routing.

\begin{figure}
    \centering
    \includegraphics[width=1\linewidth]{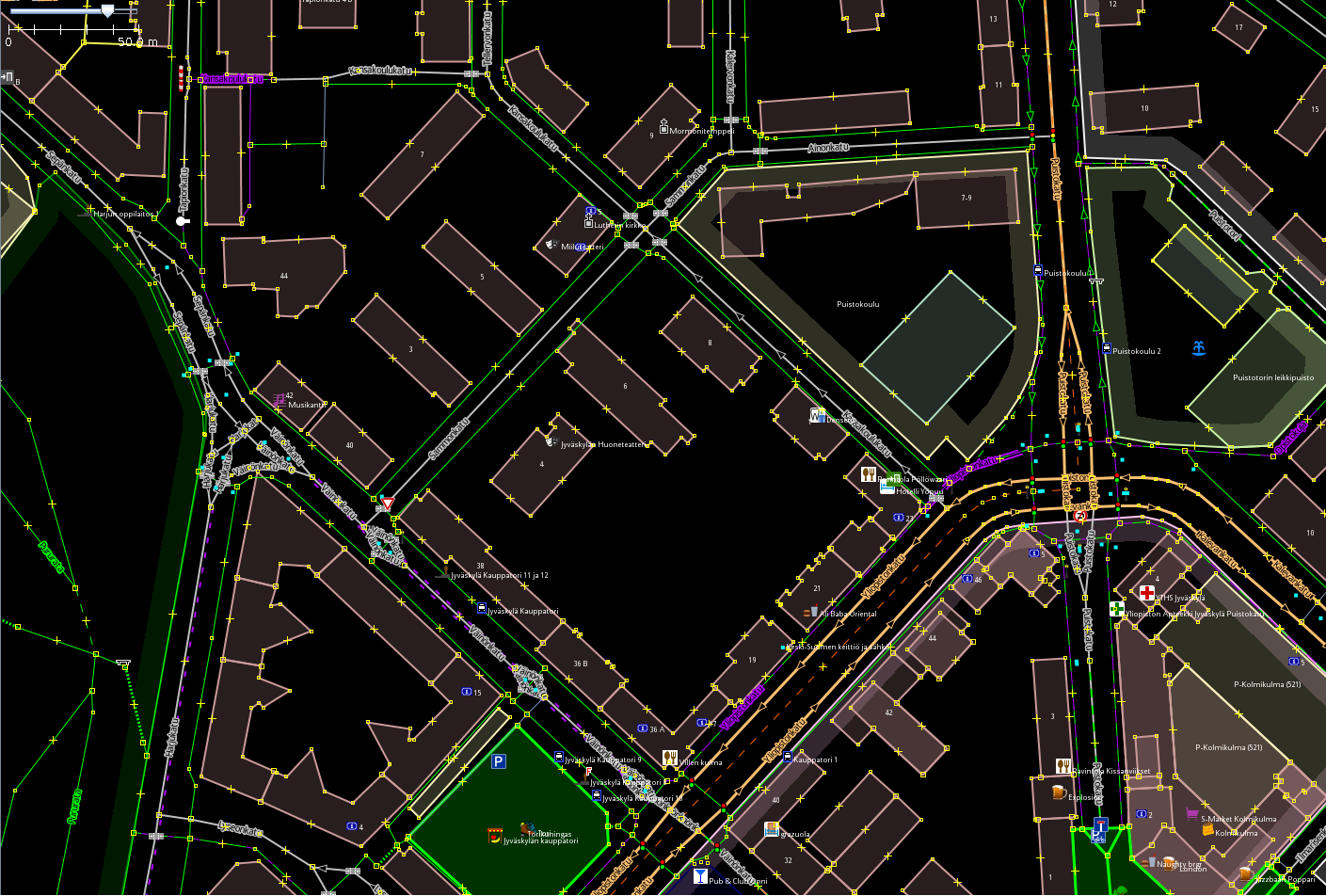}
    \caption{A part of the center of Jyv\"askyl\"a city, Finland viewed as an \emph{OSM} file in Java OpenStreetMap Editor.} 
    \label{fig:josm_default00}
\end{figure}

\begin{figure}
    \centering
    \includegraphics[width=1\linewidth]{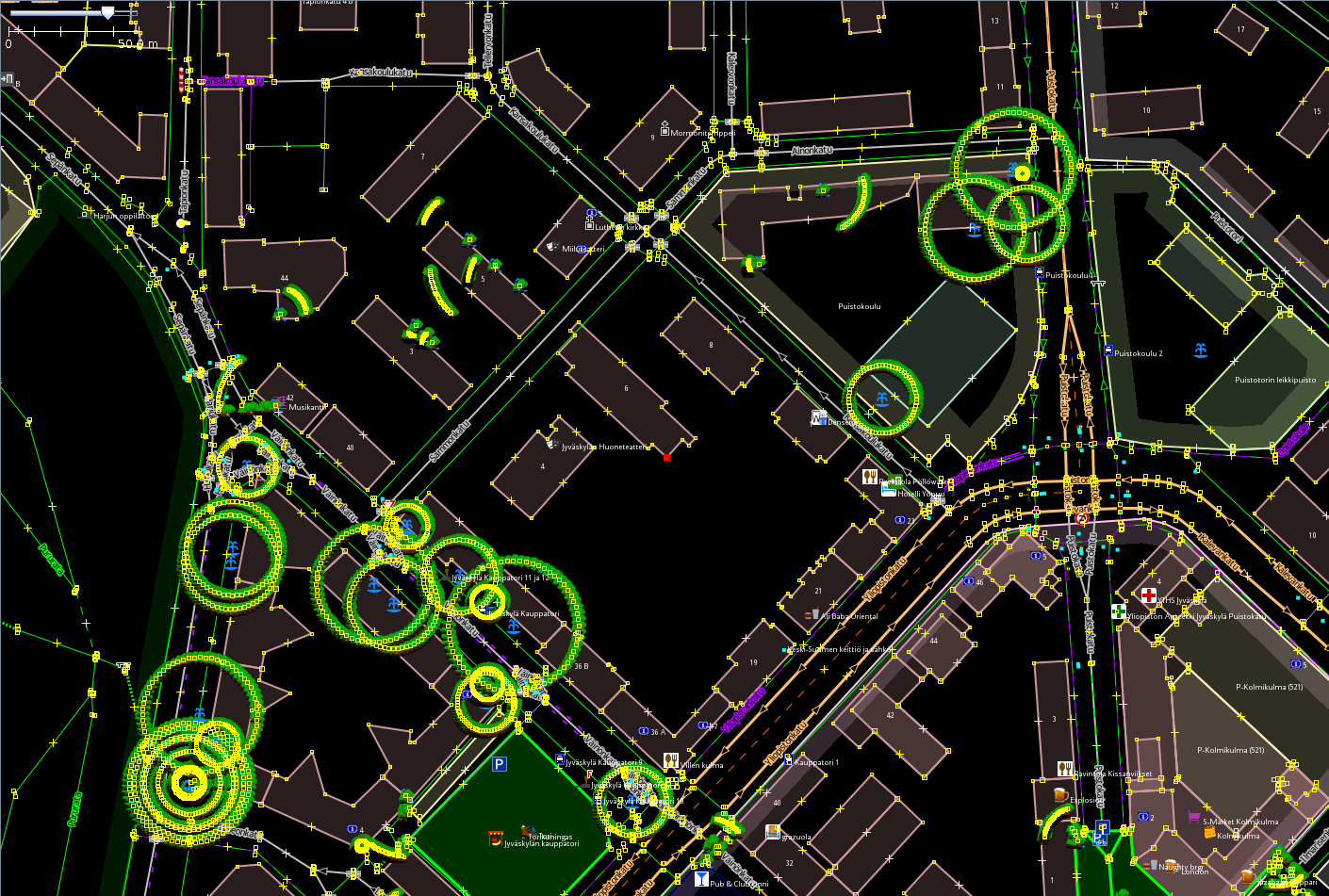}
    \caption{A part of the center of Jyv\"askyl\"a city, Finland, viewed as an \emph{OSM} file in Java OpenStreetMap Editor with CCTV cameras added with a script. The radii, angles of view and directions of the cameras have been added randomly.} 
    \label{fig:josm_default_cams00}
\end{figure}

Figure~\ref{fig:josm_default00} and Figure~\ref{fig:josm_default_cams00} show an area from the center of Jyv\"askyl\"a city, Finland. In Figure~\ref{fig:josm_default00}, a normal \emph{OSM} file is examined. In Figure~\ref{fig:josm_default_cams00}, we have used our script to add CCTV cameras to locations based on data we have gathered during our earlier feasibility-study work~\cite{lahtinen2021towards}. Fields of vision of cameras tagged as \emph{round} can be seen as yellow circles that consist of \emph{OSM} nodes.  Here, the camera itself is illustrated with a rather randomly chosen figure, a fountain, in the center of the circle. Furthermore, yellow arches in varying sizes represent the fields of vision of cameras tagged as \emph{directed}. 
For evaluation purposes, in both of \emph{round} and \emph{directed} cameras, the radii, the angles of view and the directions have been given random but realistic values, as our current data does not always include the real values for them. On the other hand, such a random value assignment script is also handy to quickly generate synthetic examples, and spot any bugs or inconsistencies in our routing engine and algorithms. 

\begin{figure}
    \centering
    \includegraphics[width=1\linewidth]{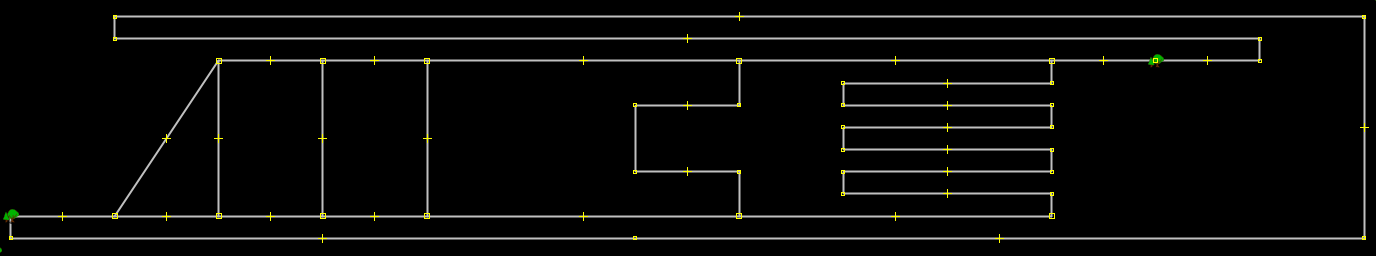}
    \caption{A plain evaluation map before any data manipulation.}
    \label{fig:josm_00}
\end{figure}

\begin{figure}
    \centering
    \includegraphics[width=1\linewidth]{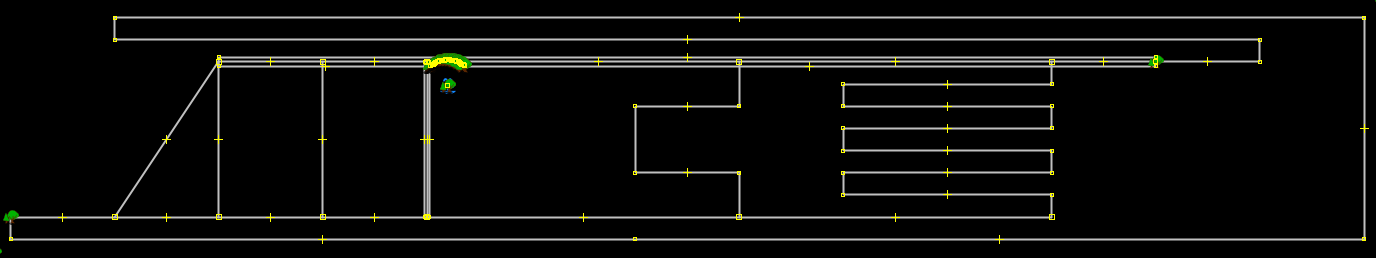}
    \caption{An evaluation map with one directed CCTV camera and split ways nearby it.} 
    \label{fig:josm_privacy04}
\end{figure}

\begin{figure}
    \centering
    \includegraphics[width=1\linewidth]{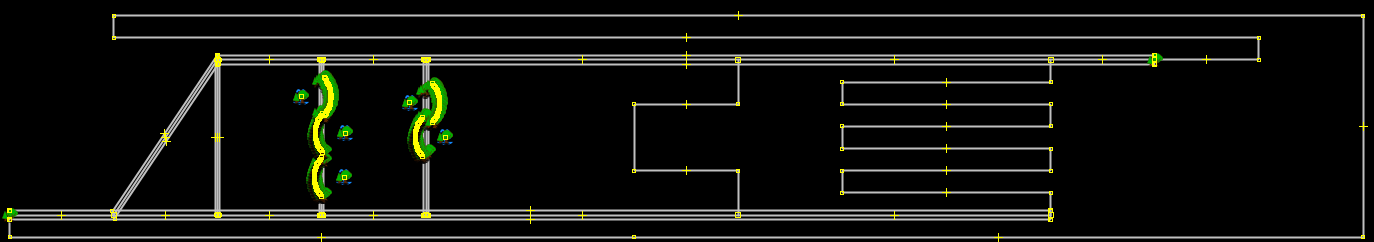}
    \caption{An evaluation map with multiple directed CCTV cameras and split ways nearby them.} 
    \label{fig:josm_safety03}
\end{figure}

Figure~\ref{fig:josm_00} is an image of the synthetic evaluation map used for the case examples. This image illustrates the default state of the evaluation map before any data manipulation has been executed. In Figure~\ref{fig:josm_privacy04}, there is a \emph{directed} CCTV camera added by a script. The ways near the camera are split in three distinct ways, all of which can be routed through. Figure~\ref{fig:josm_safety03} illustrates a similar case with more CCTV cameras added to the data.

\begin{figure}
    \centering
    \includegraphics[width=1\linewidth]{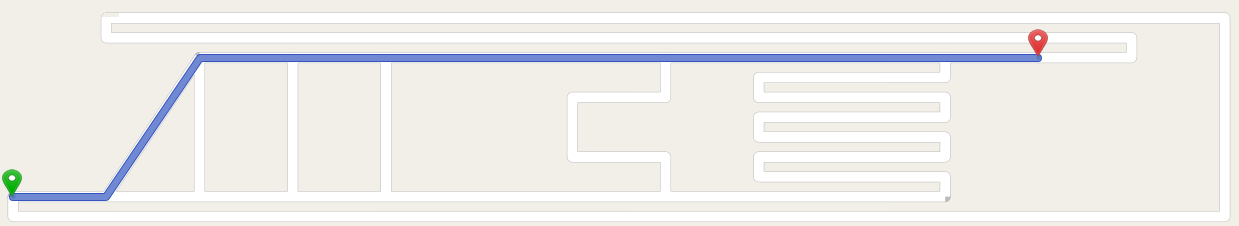}
    \caption{Default behavior of the standard \emph{OSRM}. In this case, the shortest route is used.}
    \label{fig:00}
\end{figure}

Next, before proceeding to either \emph{privacy-mode} or \emph{safety-mode}, Figure~\ref{fig:00} demonstrates the default behavior of the standard \emph{OSRM} in its frontend. In this case, the shortest possible route is being used.

\subsection{Privacy-mode}

In this section, to demonstrate how the \emph{privacy-mode} of our \emph{CCTV-aware} solution operates, our evaluations are presented as examples of some relevant cases. \emph{Privacy-mode} aims to maintain privacy by avoiding surveilled areas. It operates in a strict way as it never routes through a field of vision of a CCTV camera. If only a part of a way is under surveillance, the route goes from the other side of that way. Here, the other side means one of the split ways parallel to the original one in the middle. 

\begin{figure}
    \centering
    \includegraphics[width=1\linewidth]{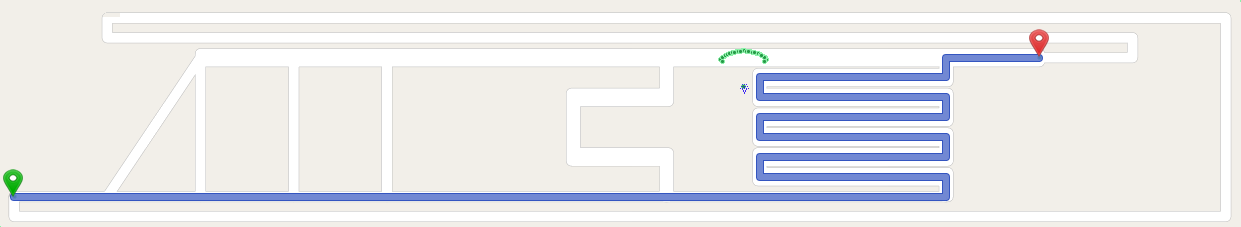}
    \caption{All the shortest ways are blocked. \emph{OSRM-CCTV} chooses the shortest route without surveillance.}
    \label{fig:privacy_01}
\end{figure}

\begin{figure}
    \centering
    \includegraphics[width=1\linewidth]{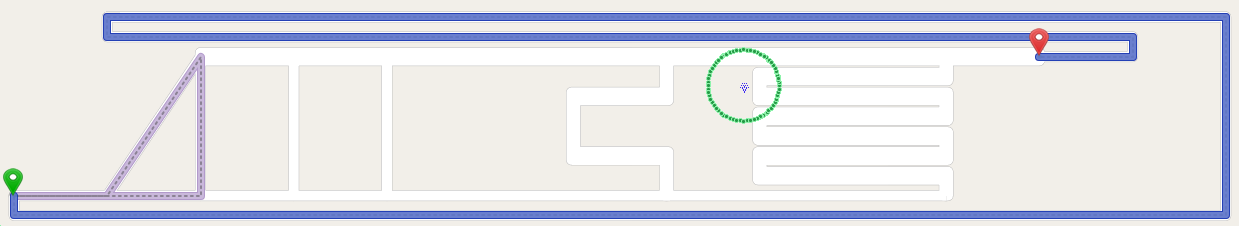}
    \caption{Even the previous route in Figure~\ref{fig:privacy_01} is closed. The longest possible route must be used.}
    \label{fig:privacy_02}
\end{figure}

\begin{figure}
    \centering
    \includegraphics[width=1\linewidth]{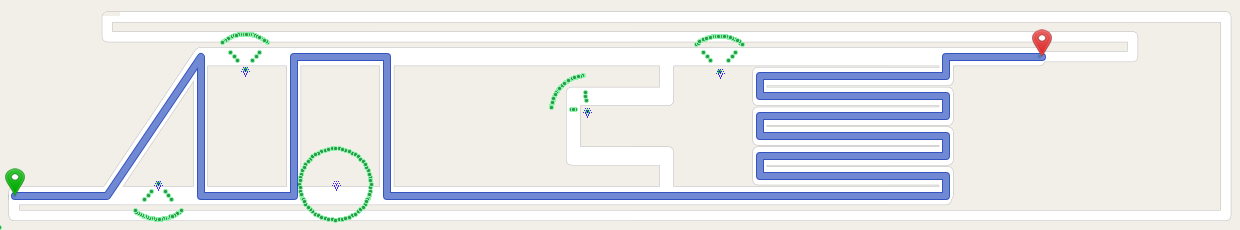}
    \caption{A more complicated situation where various parts of the map under surveillance.}
    \label{fig:privacy_021}
\end{figure}

First, in Figure~\ref{fig:privacy_01} there is an example of all the shortest ways being blocked by a CCTV camera. The calculated route will be the shortest unsurveilled one that \emph{OSRM-CCTV} can find. In Figure~\ref{fig:privacy_02}, even the route used in the previous image is blocked (see Figure~\ref{fig:privacy_01}) and the longest way on the map is used. Figure~\ref{fig:privacy_021} illustrates a more complicated route where many CCTV cameras surveil different parts of the map but there is still an unsurveilled route that is shorter than the longest possible route.

\begin{figure}
    \centering
    \includegraphics[width=1\linewidth]{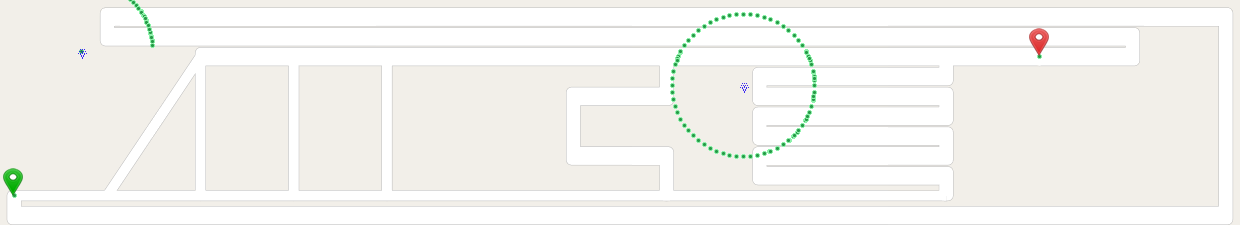}
    \caption{All routes covered by CCTV surveillance, hence \emph{OSRM-CCTV} does not compromise privacy by routing through surveillance areas and no route is generated.}
    \label{fig:privacy_03}
\end{figure}

\begin{figure}
    \centering
    \includegraphics[width=1\linewidth]{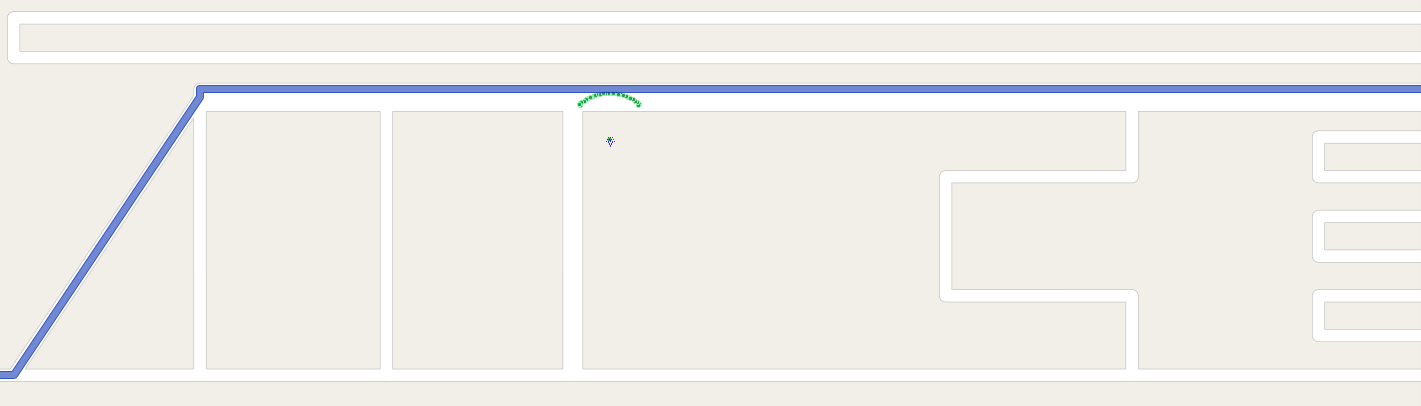}
    \caption{The area of vision of the CCTV camera reaches over only one side of the way, hence \emph{OSRM-CCTV} can still route while maintaining privacy.}
    \label{fig:privacy_04}
\end{figure}

\begin{figure}
    \centering
    \includegraphics[width=1\linewidth]{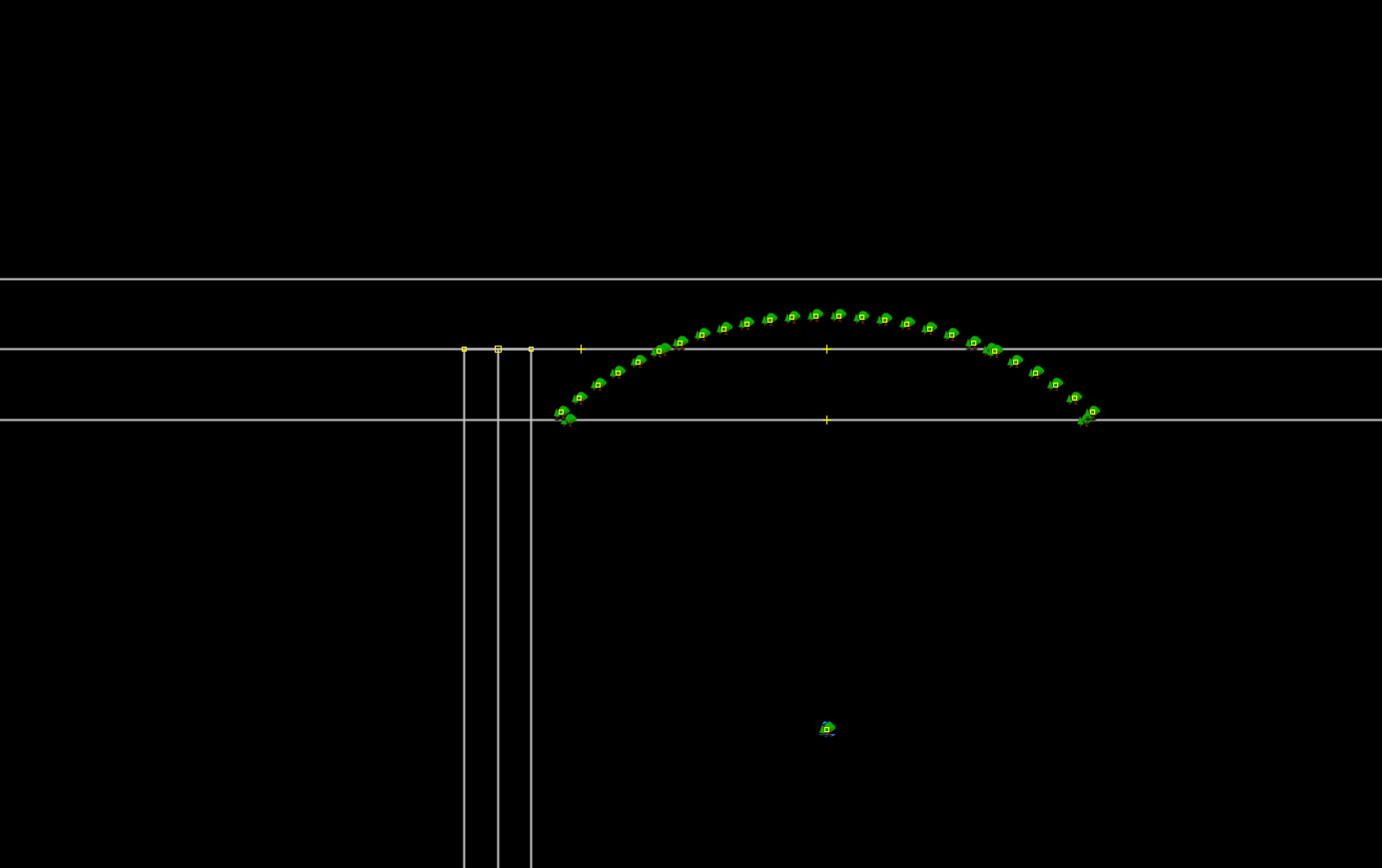}
    \caption{A zoomed-in portion of the case in Figure~\ref{fig:privacy_04} seen in \emph{JOSM}.} 
    \label{fig:josm_privacy_041}
\end{figure}

In Figure~\ref{fig:privacy_03}, there is a case where every possible route is blocked by surveillance -- no route is being generated and privacy is preserved. If only a part of a way is under surveillance (see Figure~\ref{fig:privacy_04}), \emph{privacy-mode} allows routing through the other side of the road. This is the result of using split ways to represent the dimensions of ways. Figure~\ref{fig:josm_privacy_041} shows a zoomed-in portion of area partly under surveillance where \emph{OSRM-CCTV} routes through the split way on the northern side of the road (see Figure~\ref{fig:privacy_04}).

\subsection{Safety-mode}

In this section, the \emph{safety-mode} of our \emph{CCTV-aware} solution is evaluated. The results are presented with similar case examples than earlier in the case of \emph{privacy-mode}. A safe route in \emph{safety-mode} means a route that is under the surveillance of CCTV cameras. In principle, a route leading through the fields of vision of more cameras would mean a safer route.  \emph{Safety-mode} is not simply the \emph{privacy-mode} reversed and it does not always choose a surveilled route, but it works in a more selective manner. The standard \emph{OSRM}~\cite{luxen-vetter-2011, osrm} provides traffic updates~\cite{osrm-traffic} which are used to weigh preferable routes based on the amount of CCTV cameras surveilling the way. If a route would be too long, the weighing can cause \emph{OSRM-CCTV} to generate a route that has no camera surveillance.

\begin{figure}
    \centering
    \includegraphics[width=1\linewidth]{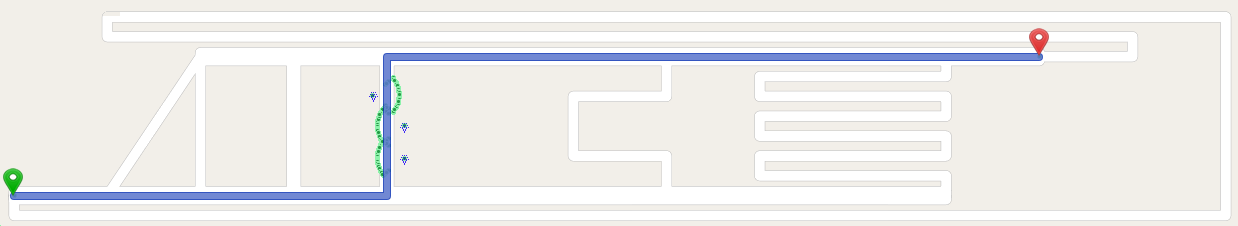}
    \caption{\emph{OSRM-CCTV} routes through the surveilled road.} 
    \label{fig:safety_01}
\end{figure}

\begin{figure}
    \centering
    \includegraphics[width=1\linewidth]{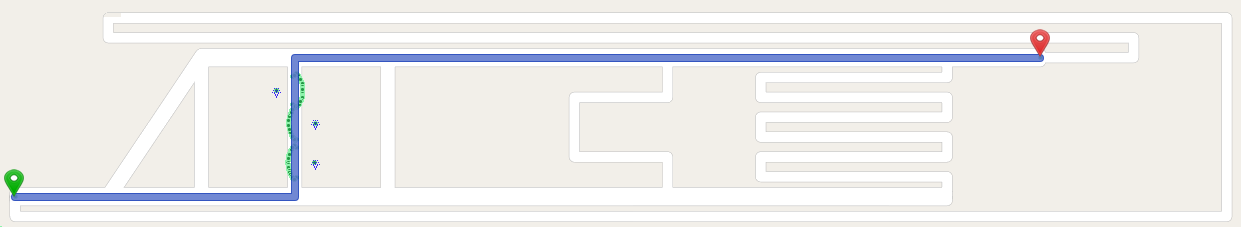}
    \caption{Similar to Figure~\ref{fig:safety_01}, \emph{OSRM-CCTV} routes through the surveilled road.} 
    \label{fig:safety_02}
\end{figure}

\begin{figure}
    \centering
    \includegraphics[width=1\linewidth]{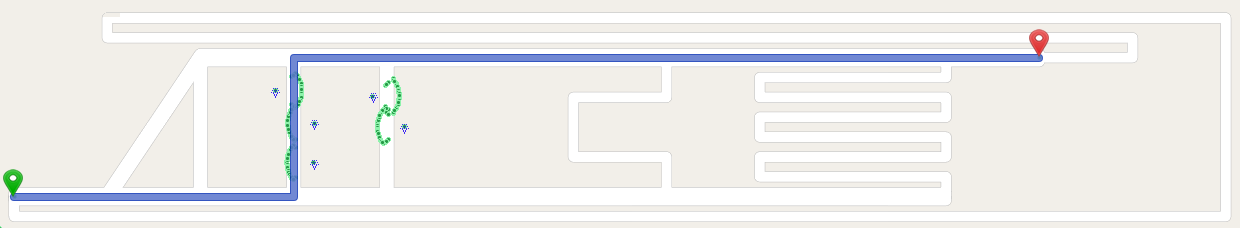}
    \caption{\emph{OSRM-CCTV} routes through the way under the surveillance of more CCTV cameras.}
    \label{fig:safety_03}
\end{figure}

\begin{figure}
    \centering
    \includegraphics[width=1\linewidth]{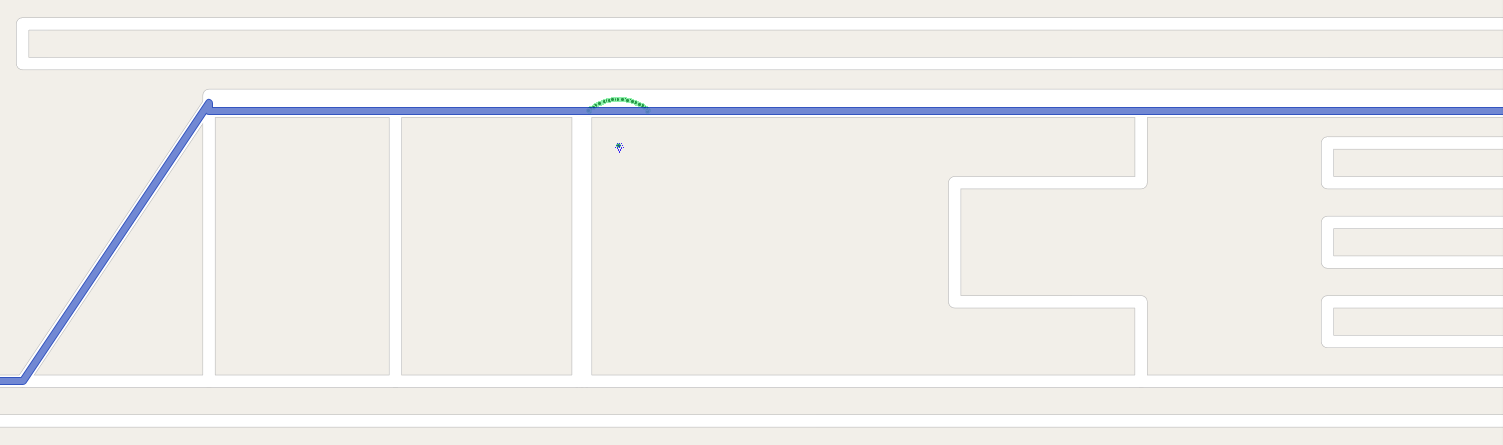}
    \caption{The small nudge after the diagonal route indicates that \emph{OSRM-CCTV} prefers routing through a longer surveilled part of the way than the other options.}
    \label{fig:safety_05}
\end{figure}

Figure~\ref{fig:safety_01} demonstrates a simple example where one of the ways is under the surveillance of three CCTV cameras. \emph{OSRM-CCTV)} chooses that way based on the weighing performed according to the \emph{OSRM-CCTV} traffic updates~\cite{osrm-traffic}. In Figure~\ref{fig:safety_02}, a similar case is displayed to prove that the chosen route is the surveilled one. In Figure~\ref{fig:safety_03}, one of the ways is surveilled by one more cameras than the next one. \emph{OSRM-CCTV} chooses the route that has more CCTV cameras surveilling it. Following Figure~\ref{fig:safety_03}, Figure~\ref{fig:safety_04} shows a similar case to prove that the chosen route is the one under the surveillance of more CCTV cameras. Figure~\ref{fig:safety_05} demonstrates the same case seen earlier with \emph{privacy-mode} where only part of a way is under surveillance (see Figure~\ref{fig:privacy_04} in last section, which is clarified with Figure~\ref{fig:josm_privacy_041}). In this case, the difference is that instead of using \emph{privacy-mode}, \emph{safety-mode} is being used. The small nudge after the diagonal part of the way indicates \emph{OSRM-CCTV} routing to a split route which leads through a longer surveilled part of the way than the other options.

\begin{figure}
    \centering
    \includegraphics[width=1\linewidth]{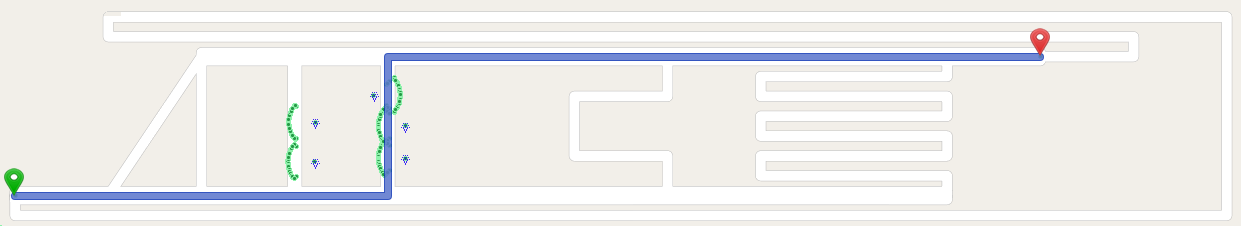}
    \caption{Similar to Figure~\ref{fig:safety_03}, \emph{OSRM-CCTV} routes through the way under the surveillance of more CCTV cameras.}
    \label{fig:safety_04}
\end{figure}

\begin{figure}
    \centering
    \includegraphics[width=1\linewidth]{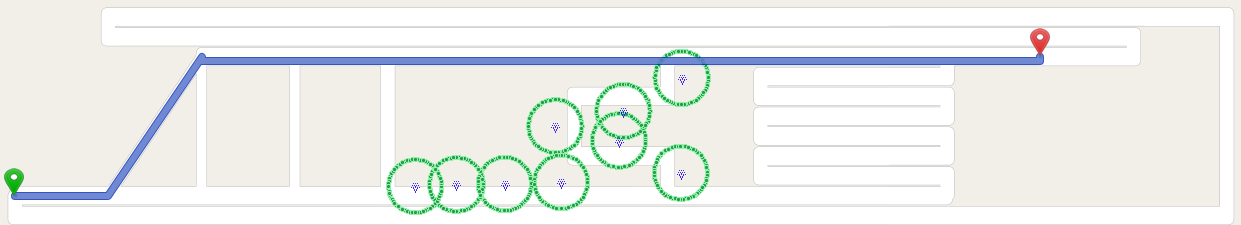}
    \caption{As intended, \emph{OSRM-CCTV} does not weigh too long a route over a safe route.} 
    \label{fig:safety_06}
\end{figure}

Lastly, as shown earlier, Figure~\ref{fig:safety_03} and Figure~\ref{fig:safety_04} demonstrate that traffic updates allow \emph{OSRM-CCTV} to prefer a surveilled route over a short one, and that \emph{OSRM-CCTV} chooses a route that is under the surveillance of more CCTV cameras.  Figure~\ref{fig:safety_06} shows \emph{OSRM-CCTV} choosing the shortest route instead of following too long a route. To some extent, this is intended behavior since one would not likely prefer to use an arbitrarily long route only to reach as many CCTV cameras as possible. However, in our evaluations we could not determine exactly how \emph{OSRM-CCTV} with its traffic updates calculates the preference between the route length and the weights of the route. For example, choosing an arbitrarily large rate value (see standard \emph{OSRM} Traffic~\cite{osrm-traffic}) one would expect \emph{OSRM-CCTV} to route through a correspondingly long route as long as it has many CCTV cameras surveilling it. As that does not seem to be the case, fine-tuning the \emph{safety-mode} would require more evaluation.

\section{Real-world examples}
\label{sec:realworld}

In this section, we present five examples of how \emph{privacy-mode} and \emph{safety-mode} of our \emph{CCTV-aware} approach operate in the downtown of the city of Jyv\"askyl\"a, Finland. This location was chosen because we have mapped around 450 CCTV cameras around the city of Jyv\"askyl\"a~\cite{lahtinen2021towards}. The mapped data contains for example the coordinates and the type of the cameras. The type can be either \emph{round} or \emph{directed}. The \emph{round} type corresponds to a camera with a circular, 360 degree angle of view. The \emph{directed} type corresponds to a camera with a less than 360 degree angle of view. In our current data and for the purposes of our evaluations, the radii, the angles of view and the directions have been added randomly with a script. 

Three of the five examples are depicted as a pair of images, both of which have the same starting and ending point. The starting point is marked with a green marker and the endpoint is marked with a red marker. Regarding the first three examples, the difference between the images in each pair is that the first one operates on \emph{privacy-mode} and the second one on \emph{safety-mode}. The last two of the five examples compare the default mode of \emph{OSRM-CCTV} routing and the \emph{safety-mode} of our \emph{CCTV-aware} approach. These two are compared to demonstrate that in \emph{safety-mode}, the router will prefer the longer route over the default route that is shorter, as long as the calculated route is not too long.

\begin{figure}
    \centering
    \includegraphics[width=0.9\linewidth]{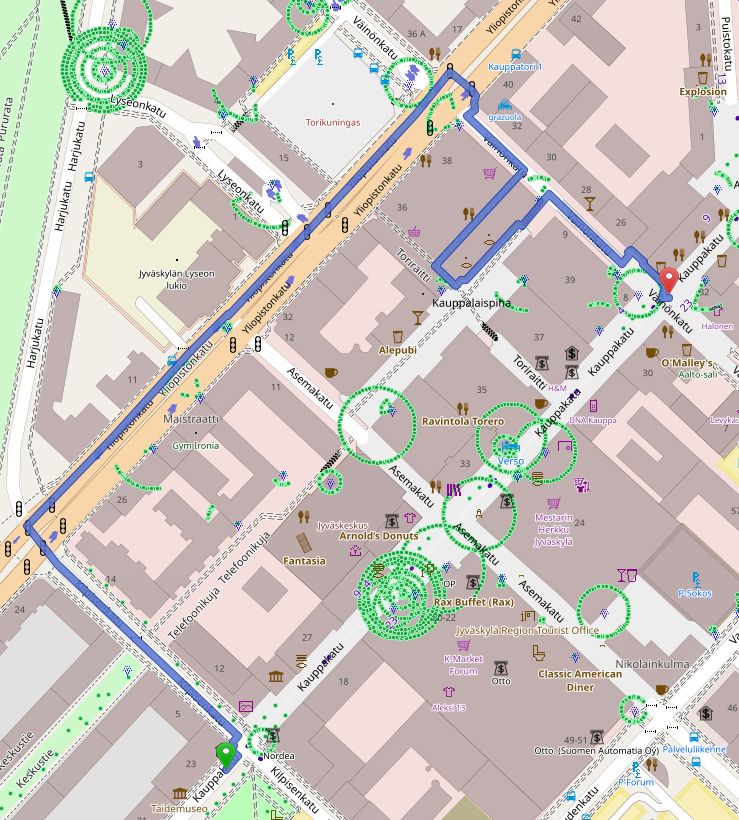}
    \caption{\emph{OSRM-CCTV} run with \emph{privacy-mode} avoids all CCTV cameras in the downtown of Jyv\"askyl\"a, Finland.} 
    \label{fig:real_privacy_01}
\end{figure}

\begin{figure}
    \centering
    \includegraphics[width=0.9\linewidth]{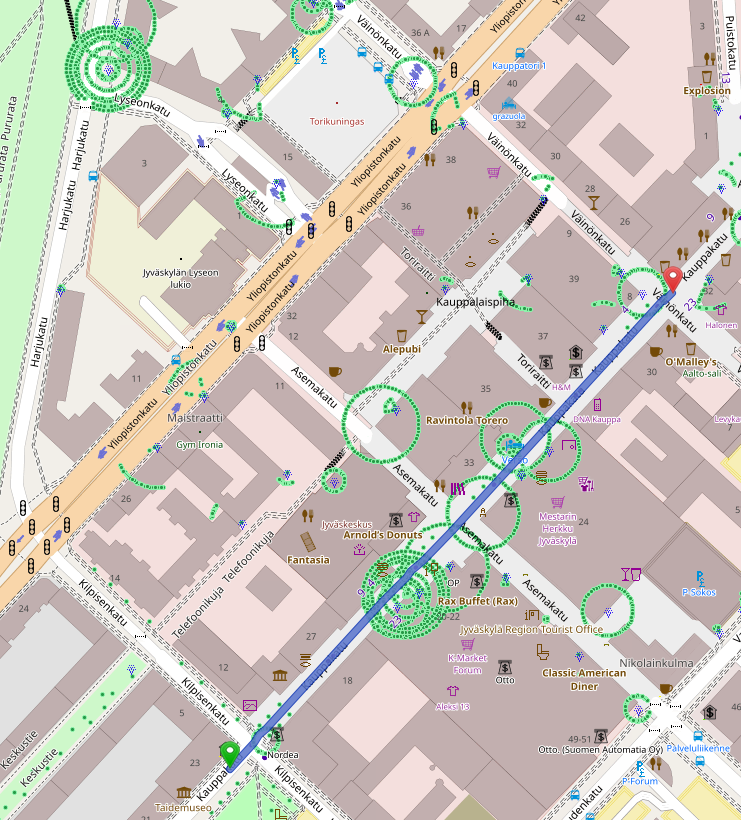}
    \caption{\emph{OSRM-CCTV} run with \emph{safety-mode} routes straight through the fields of vision of multiple CCTV cameras in the downtown of Jyv\"askyl\"a, Finland.} 
    \label{fig:real_safety_01}
\end{figure}

In the first example, Figure~\ref{fig:real_privacy_01} shows a route through the middle of the downtown of the city of Jyv\"askyl\"a. \emph{Privacy-mode} is being used, and \emph{OSRM-CCTV} does not route through the field of vision of any CCTV cameras. Figure~\ref{fig:real_safety_01} shows the same scenario run with \emph{safety-mode}. The route goes straight through the fields of vision of multiple CCTV cameras.

\begin{figure}
    \centering
    \includegraphics[width=0.9\linewidth]{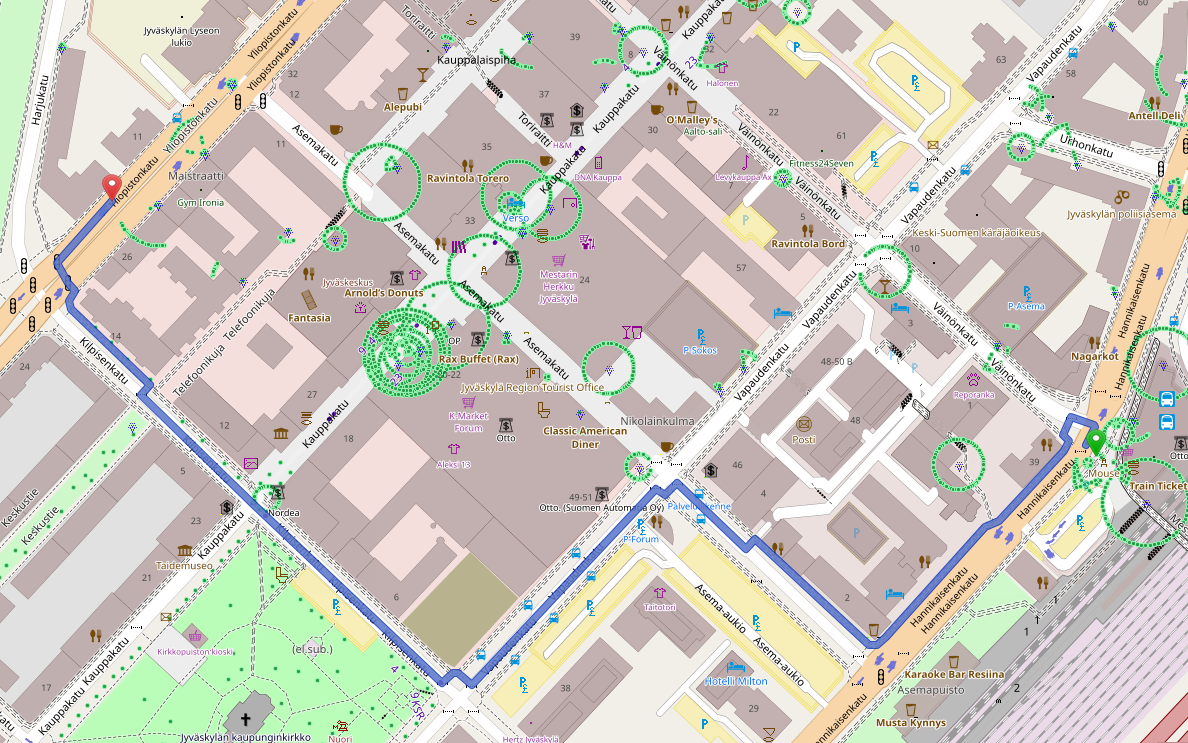}
    \caption{\emph{OSRM-CCTV} run with \emph{privacy-mode} avoids all CCTV cameras in the downtown of Jyv\"askyl\"a, Finland. Near the green marker on the right side of the image, \emph{OSRM-CCTV} switches to the opposite side of the road to avoid nearby CCTV cameras.} 
    \label{fig:real_privacy_02}
\end{figure}

\begin{figure}
    \centering
    \includegraphics[width=0.9\linewidth]{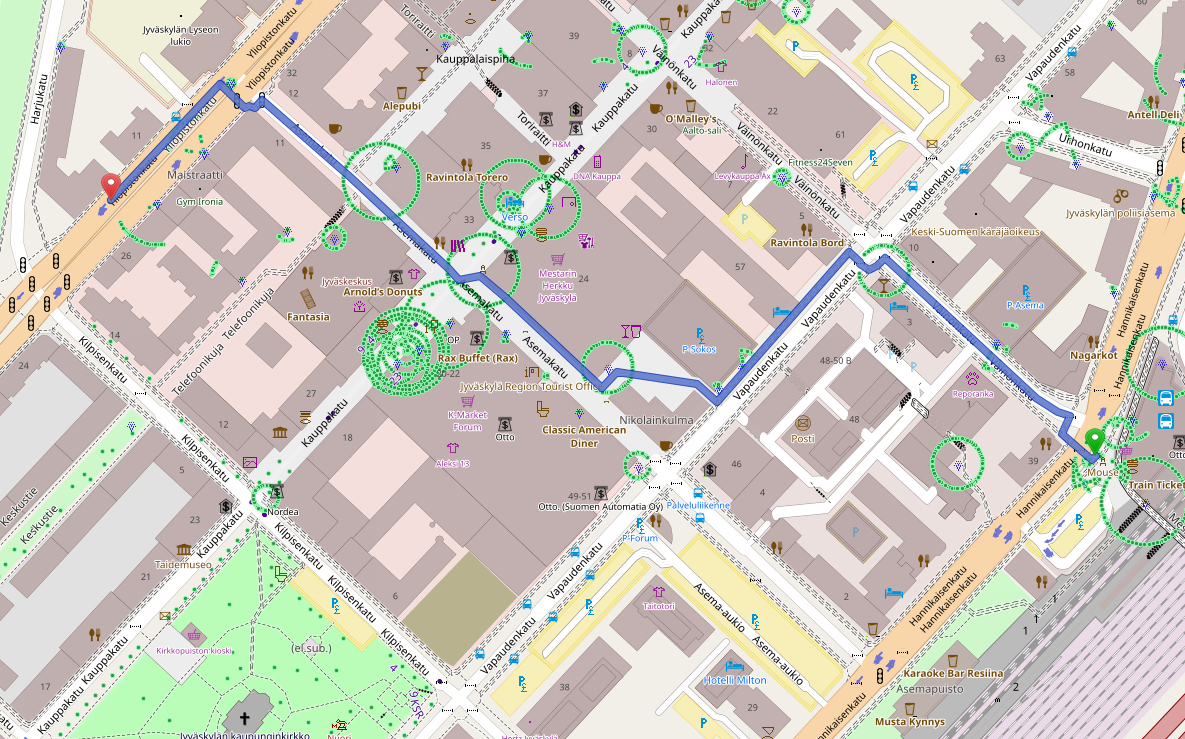}
    \caption{\emph{OSRM-CCTV} run with \emph{safety-mode} chooses the its route to route through the fields of vision of multiple CCTV cameras in the downtown of Jyv\"askyl\"a, Finland.} 
    \label{fig:real_safety_02}
\end{figure}

In the second example, \emph{privacy-mode} is run in Figure~\ref{fig:real_privacy_02}. Starting from the green marker, \emph{OSRM-CCTV} first avoids the CCTV cameras right at the start of the route by switching to the other side of the route. After that the route leads to the endpoint marked with a red marker, avoiding all possible CCTV cameras. In Figure~\ref{fig:real_safety_02}, \emph{OSRM-CCTV} operates on \emph{safety-mode} and the route goes through the middle of the map and many CCTV cameras.

\begin{figure}
    \centering
    \includegraphics[width=0.9\linewidth]{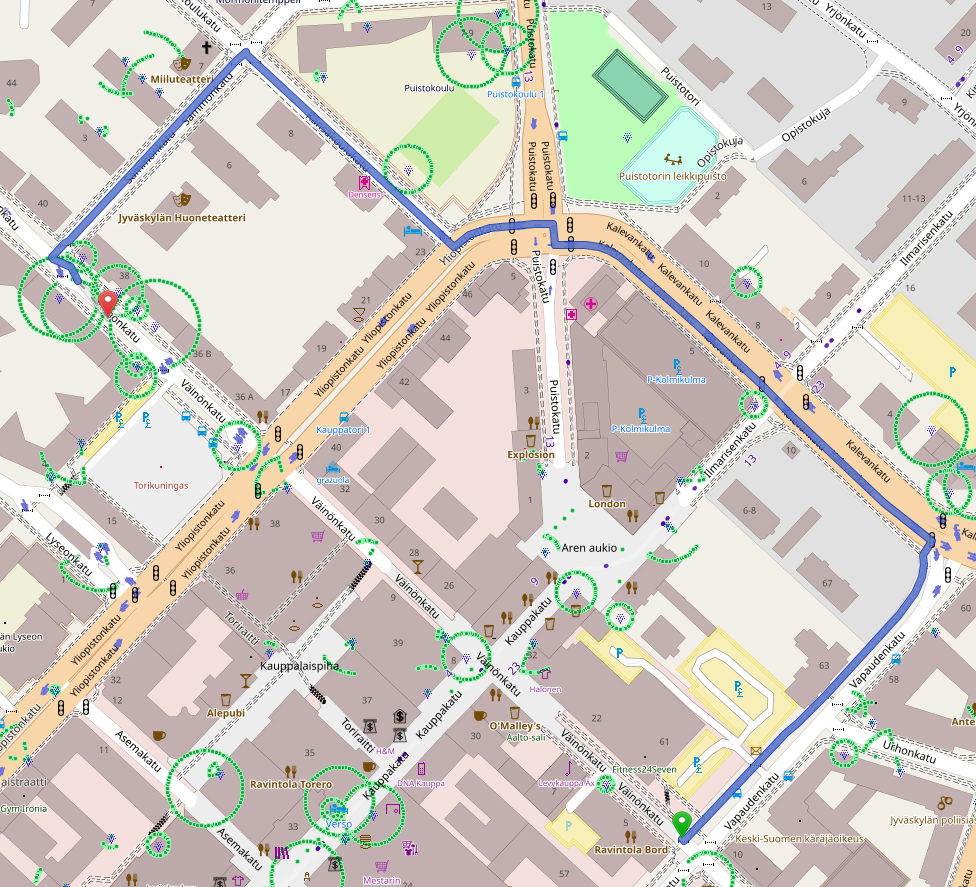}
    \caption{\emph{OSRM-CCTV} run with \emph{privacy-mode} avoids all possible CCTV cameras but does not route to the endpoint on the left side of the image. Avoiding fields of vision of CCTV cameras would be impossible there, hence privacy is maintained and the endpoint is reached only as closely as possible.}
    \label{fig:real_privacy_03}
\end{figure}

\begin{figure}
    \centering
    \includegraphics[width=0.9\linewidth]{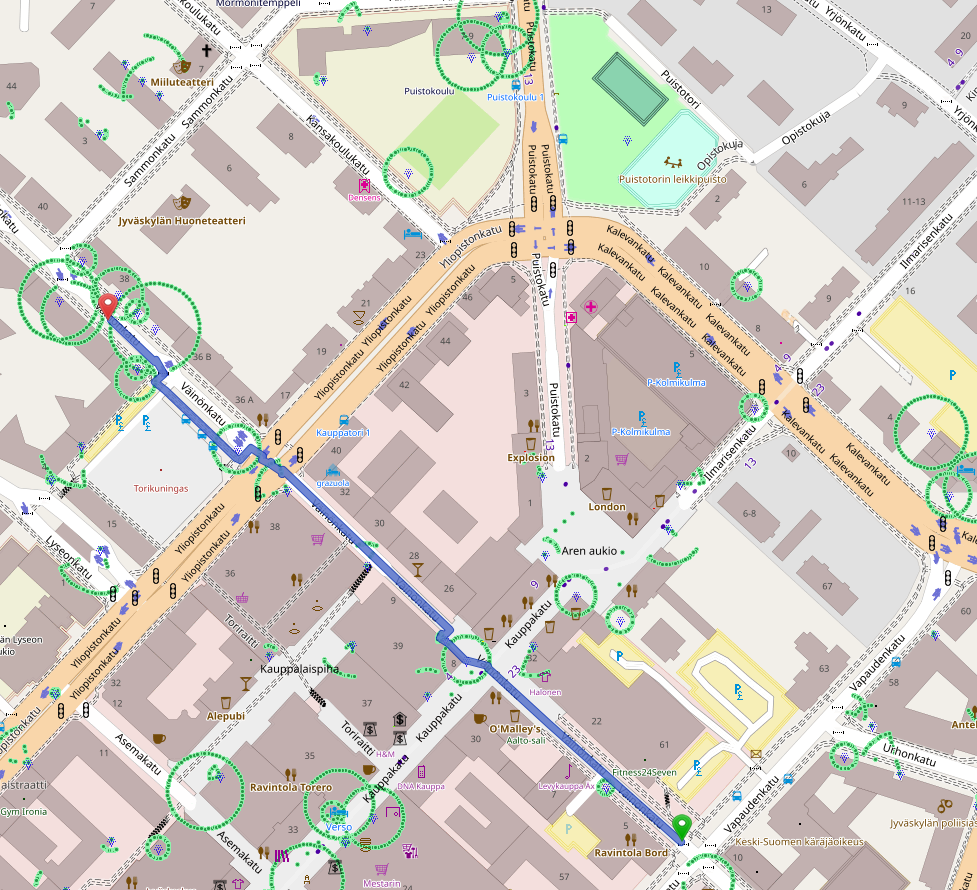}
    \caption{\emph{OSRM-CCTV} run with \emph{safety-mode} chooses the its route to route through the fields of vision of multiple CCTV cameras in the downtown of Jyv\"askyl\"a, Finland, making some turns to reach as many fields of vision of CCTV cameras as possible.} 
    \label{fig:real_safety_03}
\end{figure}

The third example shows a situation that is different than the earlier two.  Figure~\ref{fig:real_privacy_03} demonstrates how running \emph{OSRM-CCTV} in \emph{privacy-mode} avoids all possible cameras until the end of the route. Since the endpoint is in between of multiple CCTV cameras and it is impossible to route there without travelling through the field of vision of a CCTV camera, \emph{OSRM-CCTV} routes as close as it can but no further. Privacy is not compromised as the route does not go through the field of vision of any CCTV cameras. For comparison, running \emph{safety-mode} in Figure~\ref{fig:real_safety_03} the route goes from the starting point straight to the endpoint, making some turns to reach as many fields of vision of CCTV cameras as possible.

\begin{figure}
    \centering
    \includegraphics[width=0.9\linewidth]{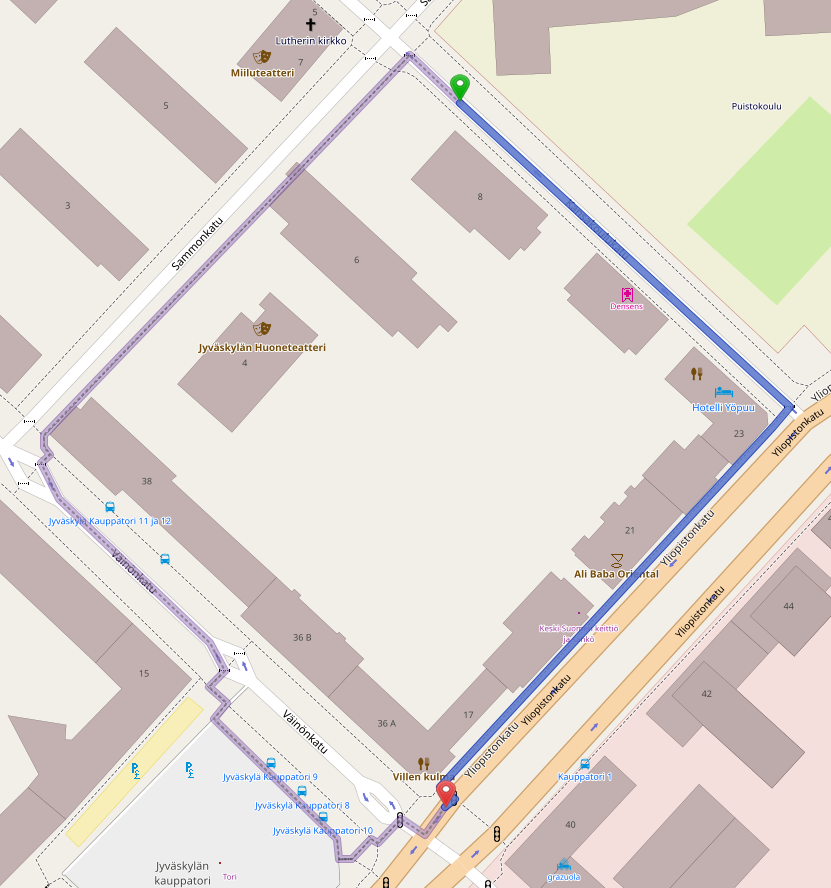}
    \caption{Running on the default routing mode, \emph{OSRM} chooses the shortest route.}
    \label{fig:real_cams_no1}
\end{figure}

\begin{figure}
    \centering
    \includegraphics[width=0.9\linewidth]{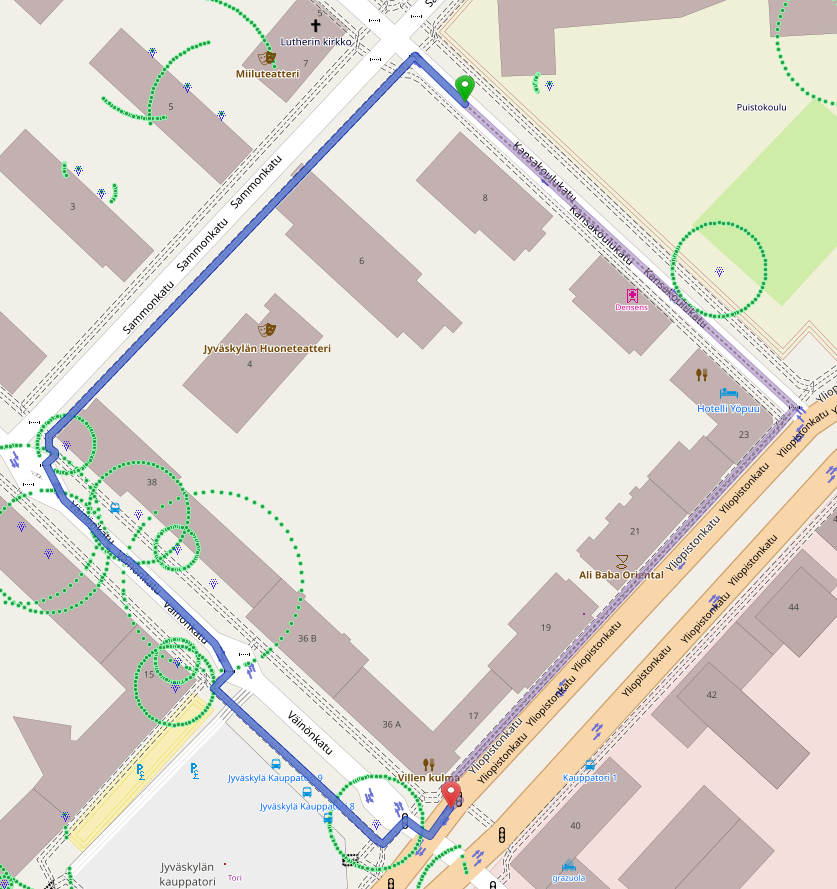}
    \caption{Running on the \emph{safety-mode}, \emph{OSRM-CCTV} chooses the longer route because the calculations are weighed by the areas of vision of multiple CCTV cameras.}
    \label{fig:real_cams_yes1}
\end{figure}

Next, the last two examples are demonstrated between the default mode of \emph{OSRM-CCTV} routing and the \emph{safety-mode} of our \emph{CCTV-aware} approach. These are compared to show that in \emph{safety-mode} the router will prefer the longer route over the default, shorter route. In Figure~\ref{fig:real_cams_no1}, when the default mode of \emph{OSRM-CCTV} routing is used, there is a situation where \emph{OSRM-CCTV} chooses the shortest, 277-meter route by default instead of the longer, 351-meter alternative route seen in shaded purple color. When \emph{safety-mode} is on (see Figure~\ref{fig:real_cams_yes1}), conversely, \emph{OSRM-CCTV} chooses the longer route that leads through the areas of vision of multiple CCTV cameras.

\begin{figure}
    \centering
    \includegraphics[width=0.9\linewidth]{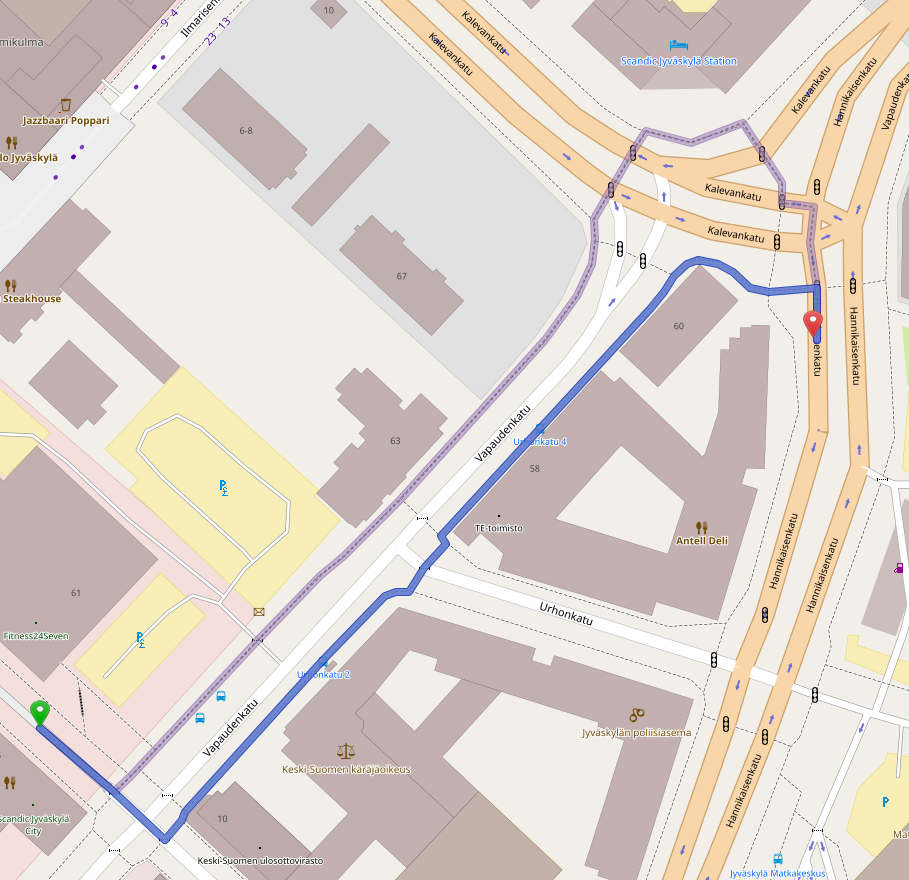}
    \caption{Running on the default routing mode, \emph{OSRM} chooses the shortest route.}
    \label{fig:real_cams_no2}
\end{figure}

\begin{figure}
    \centering
    \includegraphics[width=0.9\linewidth]{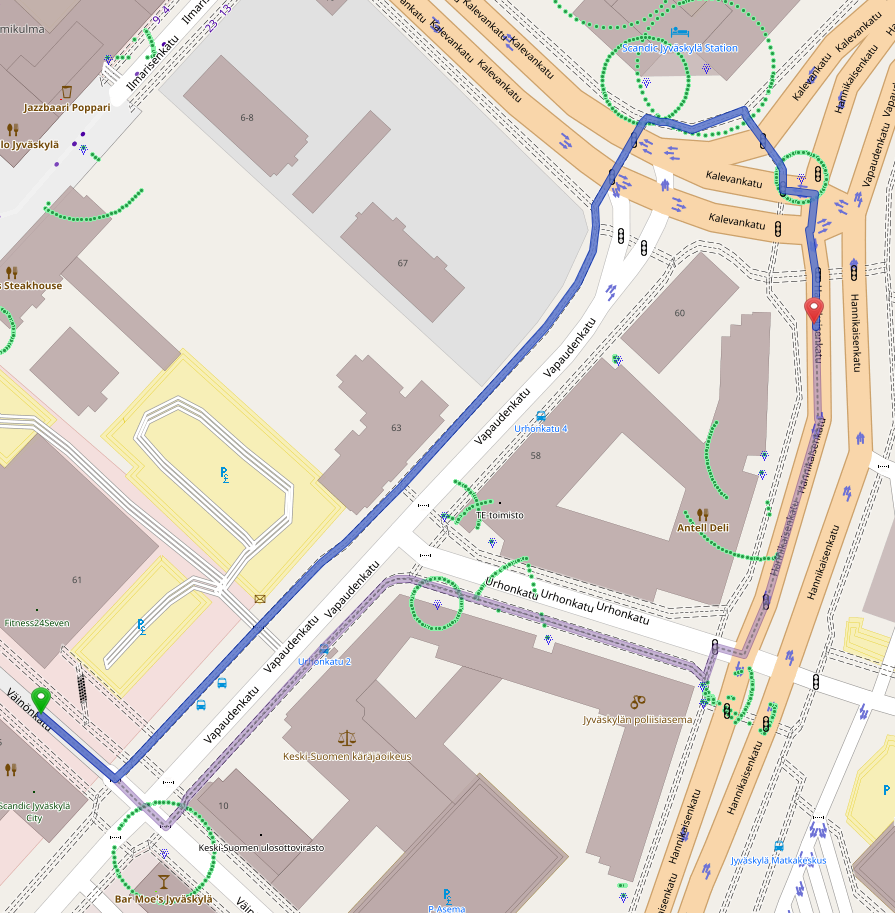}
    \caption{Running on the \emph{safety-mode}, \emph{OSRM-CCTV} chooses the longer route because the calculations are weighed by the areas of vision of multiple CCTV cameras.}
    \label{fig:real_cams_yes2}
\end{figure}

Finally, the last example is similar to the previous one. When running on \emph{safety-mode}, \emph{OSRM-CCTV} chooses the longer, 366-meter route (see Figure~\ref{fig:real_cams_yes2}) instead of the shorter 320-meter route that is chosen by default (see Figure~\ref{fig:real_cams_no2}). The alternative route seen in shaded purple color is different than in the case without CCTV cameras (see Figure~\ref{fig:real_cams_no2}) because also the alternative route leads through the fields of vision of CCTV cameras. The chosen route is weighed to be more preferable than the alternative one.

\section{Conclusion}
\label{sec:concl}

In this paper, we propose, implement and evaluate a \emph{CCTV-aware} privacy- and safety-oriented routing solution. To the best of our knowledge, this is the first work where a \emph{CCTV-aware} approach to routing has been conceptualized and proposed. For the routing engine, we start from existing open-source \emph{Open Street Routing Machine (OSRM)} routing framework. Subsequently we customize it, and open-source release \emph{OSRM-CCTV} -- a CCTV-aware routing modification based on \emph{OSRM}. 
We also validate and demonstrate the effectiveness and usability of the system on a handful of 
synthetic and real-world examples.
To help validate our work and to further encourage the development and wide adoption of the system, 
we release the relevant artifacts (e.g., code, data, documentation) as open-source: \url{https://github.com/Fuziih}



\section*{Acknowledgments}

We acknowledge grants of computer capacity from the 
Finnish Grid and Cloud Infrastructure (FGCI) 
(persistent identifier \texttt{urn:nbn:fi:research-infras-2016072533}).
Part of this research was kindly supported by a 
\emph{``17.06.2020 Decision of the Research Dean on research funding within the faculty''} 
grant from the Faculty of Information Technology of the University of Jyv\"askyl\"a, 
and the grant was facilitated and managed by Dr. Andrei Costin.
The authors would also like to thank the following persons for their dedicated efforts during 
the crowdsource data collection and/or internal paper review phase(s): 
Syed Khandker (special mention and many thanks).

Hannu Turtiainen would like to also thank the Finnish Cultural Foundation / Suomen Kulttuurirahasto (https://skr.fi/en) 
for supporting his PhD dissertation work and research (grant decision 00211119), and the Faculty of Information Technology of University of Jyv\"{a}skyl\"{a} (JYU), 
in particular Prof. Timo H\"{a}m\"{a}l\"{a}inen, for partly supporting his PhD supervision at JYU during 2021--2022.



\bibliographystyle{IEEEtran}
\bibliography{IEEE.bib}

\begin{thebibliography}{10}
\providecommand{\url}[1]{#1}
\csname url@samestyle\endcsname
\providecommand{\newblock}{\relax}
\providecommand{\bibinfo}[2]{#2}
\providecommand{\BIBentrySTDinterwordspacing}{\spaceskip=0pt\relax}
\providecommand{\BIBentryALTinterwordstretchfactor}{4}
\providecommand{\BIBentryALTinterwordspacing}{\spaceskip=\fontdimen2\font plus
\BIBentryALTinterwordstretchfactor\fontdimen3\font minus
  \fontdimen4\font\relax}
\providecommand{\BIBforeignlanguage}[2]{{%
\expandafter\ifx\csname l@#1\endcsname\relax
\typeout{** WARNING: IEEEtran.bst: No hyphenation pattern has been}%
\typeout{** loaded for the language `#1'. Using the pattern for}%
\typeout{** the default language instead.}%
\else
\language=\csname l@#1\endcsname
\fi
#2}}
\providecommand{\BIBdecl}{\relax}
\BIBdecl

\bibitem{bischoff-2021}
P.~Bischoff, ``Surveillance {{Camera Statistics}}: {{Which City}} has the
  {{Most CCTV Cameras}}?''
  \url{https://www.comparitech.com/blog/vpn-privacy/the-worlds-most-surveilled-cities/},
  May 2021.

\bibitem{cnbc2019billion}
E.~Cosgrove, ``{One billion surveillance cameras will be watching around the
  world in 2021},''
  \url{https://cnbc.com/2019/12/06/one-billion-surveillance-cameras-will-be-watching-globally-in-2021.html}.

\bibitem{caughtoncamera2019cctvlond}
``{How Many CCTV Cameras in London?}''
  \url{https://www.caughtoncamera.net/news/how-many-cctv-cameras-in-london/}.

\bibitem{ipvm2016us}
B.~Karas, ``{Americans Vastly Underestimate Being Recorded on CCTV},''
  \url{https://ipvm.com/reports/america-cctv-recording}.

\bibitem{bi2019cctv}
J.~Pasley, ``{I documented every surveillance camera on my way to work in New
  York City, and it revealed a dystopian reality},''
  \url{https://www.businessinsider.com/how-many-security-cameras-in-new-york-city-2019-12},
  Dec 2019.

\bibitem{vonhirsch2000ethics}
A.~von Hirsch, \emph{{The ethics of public television surveillance. Ethical and
  Social Perspectives on Situational Crime Prevention.}}\hskip 1em plus 0.5em
  minus 0.4em\relax Hart Publishing, 2000.

\bibitem{hu2004survey}
W.~Hu, T.~Tan, L.~Wang, and S.~Maybank, ``A survey on visual surveillance of
  object motion and behaviors,'' \emph{IEEE Transactions on Systems, Man, and
  Cybernetics, Part C (Applications and Reviews)}, vol.~34, no.~3, pp.
  334--352, 2004.

\bibitem{wheeler2010face}
F.~W. Wheeler, R.~L. Weiss, and P.~H. Tu, ``{Face recognition at a distance
  system for surveillance applications},'' in \emph{2010 Fourth IEEE
  International Conference on Biometrics: Theory, Applications and Systems
  (BTAS)}.\hskip 1em plus 0.5em minus 0.4em\relax IEEE, 2010, pp. 1--8.

\bibitem{axis-id-rec}
Axis, ``Identification and recognition,''
  \url{https://www.axis.com/files/feature_articles/ar_id_and_recognition_53836_en_1309_lo.pdf}.

\bibitem{larsen2011setting}
B.~v. S.-T. Larsen, \emph{{Setting the watch: privacy and the ethics of CCTV
  surveillance}}.\hskip 1em plus 0.5em minus 0.4em\relax Bloomsbury Publishing,
  2011.

\bibitem{costin2016security}
A.~Costin, ``{Security of CCTV and video surveillance systems: Threats,
  vulnerabilities, attacks, and mitigations},'' in \emph{6th International
  Workshop on Trustworthy Embedded Devices (TrustED)}, 2016.

\bibitem{osm-route-online}
{OpenStreetMap}, ``{Routing/online routers},''
  \url{https://wiki.openstreetmap.org/wiki/Routing/online_routers}.

\bibitem{osm-route-offline}
------, ``{Routing/offline routers},''
  \url{https://wiki.openstreetmap.org/wiki/Routing/offline_routers}.

\bibitem{osm-based}
{OpenStreetMap Wiki}, ``{List of OSM-based services --- OpenStreetMap Wiki},''
  \url{https://wiki.openstreetmap.org/w/index.php?title=List_of_OSM-based_services&oldid=2052956},
  2020, [Online; accessed 1-November-2020].

\bibitem{luxen2011real}
D.~Luxen and C.~Vetter, ``{Real-time routing with OpenStreetMap data},'' in
  \emph{Proceedings of the 19th ACM SIGSPATIAL international conference on
  advances in geographic information systems}, 2011.

\bibitem{bast2016route}
H.~Bast, D.~Delling, A.~Goldberg, M.~M{\"u}ller-Hannemann, T.~Pajor,
  P.~Sanders, D.~Wagner, and R.~F. Werneck, ``Route planning in transportation
  networks,'' in \emph{Algorithm engineering}.\hskip 1em plus 0.5em minus
  0.4em\relax Springer, 2016.

\bibitem{szczerba2000robust}
R.~J. Szczerba, P.~Galkowski, I.~S. Glicktein, and N.~Ternullo, ``Robust
  algorithm for real-time route planning,'' \emph{IEEE Transactions on
  Aerospace and Electronic Systems}, vol.~36, no.~3, pp. 869--878, 2000.

\bibitem{turtiainen2020towards}
H.~Turtiainen, A.~Costin, T.~Hamalainen, and T.~Lahtinen, ``{Towards
  large-scale, automated, accurate detection of CCTV camera objects using
  computer vision. Applications and implications for privacy, safety, and
  cybersecurity.(Preprint)},'' \emph{arXiv preprint arXiv:2006.03870}, 2020.

\bibitem{openrouteservice}
H.~I. f. G. T.~H. GIScience, \url{https://openrouteservice.org/}, 2020.

\bibitem{wheelchair}
\BIBentryALTinterwordspacing
P.~Kasemsuppakorn and H.~A. Karimi, ``Personalised routing for wheelchair
  navigation,'' \emph{Journal of Location Based Services}, vol.~3, no.~1, pp.
  24--54, 2009. [Online]. Available:
  \url{https://doi.org/10.1080/17489720902837936}
\BIBentrySTDinterwordspacing

\bibitem{crowdsourcing-mobility-impaired}
A.~Zipf, A.~Mobasheri, A.~Rousell, and S.~Hahmann, ``Crowdsourcing for
  individual needs—the case of routing and navigation for mobility-impaired
  persons,'' \emph{European Handbook of Crowdsourced Geographic Information},
  pp. 325--337, 2016.

\bibitem{curvature}
A.~Franco, ``What is curvature?'' \url{https://roadcurvature.com/}, 2016,
  [Online; accessed 1-November-2020].

\bibitem{shadow-as-route-2016}
C.~{Olaverri Monreal}, M.~{Pichler}, G.~{Krizek}, and S.~{Naumann}, ``{Shadow
  as Route Quality Parameter in a Pedestrian-Tailored Mobile Application},''
  \emph{IEEE Intelligent Transportation Systems Magazine}, 2016.

\bibitem{keithma2018parasol}
K.~Ma, ``{Parasol Navigation: Optimizing walking routes to keep you in the sun
  or shade},''
  \url{https://www.allnans.com/jekyll/update/2018/08/07/introducing-parasol.html},
  2018.

\bibitem{deilami2020allowing}
K.~Deilami, J.~Rudner, A.~Butt, T.~MacLeod, G.~Williams, H.~Romeijn, and
  M.~Amati, ``{Allowing Users to Benefit from Tree Shading: Using a Smartphone
  App to Allow Adaptive Route Planning during Extreme Heat},'' \emph{Forests},
  vol.~11, no.~9, p. 998, 2020.

\bibitem{quality-aware}
P.~Siriaraya, Y.~Wang, Y.~Zhang, S.~Wakamiya, P.~Jeszenszky, Y.~Kawai, and
  A.~Jatowt, ``Beyond the shortest route: A survey on quality-aware route
  navigation for pedestrians,'' \emph{IEEE Access}, vol.~PP, pp. 1--1, 07 2020.

\bibitem{parasol2018github}
K.~Ma, ``{Parasol: Shade model and routing algorithm for comfortable travel
  outdoors},'' \url{https://github.com/keithfma/parasol}.

\bibitem{osm-way}
{OpenStreetMap Wiki}, ``Way --- openstreetmap wiki{,},''
  \url{https://wiki.openstreetmap.org/w/index.php?title=Way&oldid=2045405},
  2020, [Online; accessed 2-November-2020].

\bibitem{osm-creaking}
M.~Lucas-Smith, ``Is the osm data model creaking?''
  \url{https://2019.stateofthemap.org/sessions/DW7WW8/}.

\bibitem{safe-lighting}
S.~{Bao}, T.~{Nitta}, K.~{Ishikawa}, M.~{Yanagisawa}, and N.~{Togawa}, ``A safe
  and comprehensive route finding method for pedestrian based on lighting and
  landmark,'' in \emph{2016 IEEE 5th Global Conference on Consumer
  Electronics}, 2016, pp. 1--5.

\bibitem{street-illumination}
H.~Miura, S.~Takeshima, N.~Matsuda, and H.~Taki, ``A study on navigation system
  for pedestrians based on street illuminations,'' in \emph{International
  Conference on Knowledge-Based and Intelligent Information and Engineering
  Systems}.\hskip 1em plus 0.5em minus 0.4em\relax Springer, 2011, pp. 49--55.

\bibitem{safety-aware-routing}
A.~Keler and J.~D. Mazimpaka, ``Safety-aware routing for motorised tourists
  based on open data and vgi,'' \emph{Journal of location Based services},
  vol.~10, no.~1, pp. 64--77, 2016.

\bibitem{customized-pedestrian-routes}
T.~Novack, Z.~Wang, and A.~Zipf, ``A system for generating customized pleasant
  pedestrian routes based on openstreetmap data,'' \emph{Sensors}, vol.~18,
  no.~11, p. 3794, 2018.

\bibitem{lahtinen2020feasibility}
T.~Lahtinen, L.~Sintonen, H.~Turtiainen, and A.~Costin, ``{Feasibility Study on
  CCTV-aware Routing and Navigation for Privacy, Anonymity, and Safety.
  Jyvaskyla -- Case-study of the First City to Benefit from CCTV-aware
  Technology. (Preprint)},'' 2020.

\bibitem{lahtinen2021brima}
T.~Lahtinen, H.~Turtiainen, and A.~Costin, ``{BRIMA: low-overhead BRowser-only
  IMage Annotation tool},'' in \emph{(To appear in) Proceedings of the IEEE
  International Conference on Image Processing}, 2021.

\bibitem{google-maps}
Google, \url{https://www.google.com/maps}.

\bibitem{yandex-maps}
Yandex, \url{https://yandex.eu/maps/}.

\bibitem{baidu-maps}
Baidu, \url{map.baidu.com}.

\bibitem{lahtinen2021towards}
T.~Lahtinen, L.~Sintonen, H.~Turtiainen, and A.~Costin, ``Towards cctv-aware
  routing and navigation for privacy, anonymity, and safety-feasibility study
  in jyv{\"a}skyl{\"a},'' in \emph{2021 28th Conference of Open Innovations
  Association (FRUCT)}.\hskip 1em plus 0.5em minus 0.4em\relax IEEE, 2021, pp.
  252--263.

\bibitem{luxen-vetter-2011}
\BIBentryALTinterwordspacing
D.~Luxen and C.~Vetter, ``{Real-time routing with OpenStreetMap data},'' in
  \emph{Proceedings of the 19th ACM SIGSPATIAL International Conference on
  Advances in Geographic Information Systems}, ser. GIS '11.\hskip 1em plus
  0.5em minus 0.4em\relax New York, NY, USA: ACM, 2011, pp. 513--516. [Online].
  Available: \url{http://doi.acm.org/10.1145/2093973.2094062}
\BIBentrySTDinterwordspacing

\bibitem{osrm-first}
D.~Luxen, ``{[OSM-dev] Announcing the immediate availability of the Open Source
  Routing Machine},''
  \url{https://lists.openstreetmap.org/pipermail/dev/2010-July/019834.html},
  2010.

\bibitem{lua-about}
R.~Ierusalimschy, W.~Celes, and L.~H. de~Figueiredo, ``About,''
  \url{https://www.lua.org/license.html}.

\bibitem{osrm-profiles}
Project-OSRM, ``Osrm profiles,''
  \url{https://github.com/Project-OSRM/osrm-backend/blob/master/docs/profiles.md},
  2020.

\bibitem{python}
G.~van Rossum, \url{https://www.python.org/}.

\bibitem{osmium-osmcode}
J.~Topf, ``{Osmium Library},'' \url{https://osmcode.org/libosmium/}.

\bibitem{pyosmium-osmcode}
S.~Hoffmann, ``Osmium python library,'' \url{https://osmcode.org/pyosmium/}.

\bibitem{shapely}
S.~Gillies, ``The shapely user manual,''
  \url{https://shapely.readthedocs.io/en/stable/manual.html}, 2020.

\bibitem{pyproj}
J.~Whitaker, \url{https://pypi.org/project/pyproj/}.

\bibitem{docker}
{Docker, Inc.}, \url{https://www.docker.com/company}.

\bibitem{ubuntu}
{Canonical Ltd.}, \url{https://ubuntu.com/about}.

\bibitem{nginx}
I.~Sysoev and I.~Nginx, \url{https://nginx.org/en/}.

\bibitem{osm-xml}
{OpenStreetMap Wiki}, ``Osm xml --- openstreetmap wiki{,},''
  \url{https://wiki.openstreetmap.org/w/index.php?title=OSM_XML&oldid=2027286},
  2020, [Online; accessed 5-November-2020].

\bibitem{osrm-traffic}
Project-OSRM, ``Traffic,''
  \url{https://github.com/Project-OSRM/osrm-backend/wiki/Traffic}, 2019.

\bibitem{osrm-backend}
D.~Luxen and C.~Vetter, \url{https://github.com/Project-OSRM/osrm-backend}.

\bibitem{osrm-frontend}
------, \url{https://github.com/Project-OSRM/osrm-frontend}.

\bibitem{osrm-frontend-fossgis}
D.~Luxen, C.~Vetter, and datendelphin,
  \url{https://github.com/fossgis-routing-server/osrm-frontend}.

\bibitem{tile-server}
A.~Overvoorde, ``openstreetmap-tile-server,''
  \url{https://github.com/Overv/openstreetmap-tile-server/}, 2019.

\bibitem{someoneelse}
A.~Townsend, ``Someoneelse-style,''
  \url{https://github.com/SomeoneElseOSM/SomeoneElse-style}.

\bibitem{josm}
I.~Scholz, D.~Stöcker, and other contributors,
  \url{https://josm.openstreetmap.de/}, 2020.

\bibitem{osrm}
{Project-OSRM}, ``{Open Source Routing Machine (OSRM) -- A modern C++ routing
  engine for shortest paths in road networks.}''
  \url{http://project-osrm.org/}.

\end{thebibliography}

\ifCLASSOPTIONcaptionsoff
  \newpage
\fi

\end{document}